\documentclass[a4paper]{article}
\usepackage[all]{xy}\usepackage[latin1]{inputenc}        
\usepackage[dvips]{graphics,graphicx}
\usepackage{amsfonts,amssymb,amsmath,color,mathrsfs, amstext}
\usepackage{amsbsy, amsopn, amscd, amsxtra, amsthm,authblk}
\usepackage{enumerate,algorithmic,algorithm}
\usepackage{upref}
\usepackage{geometry}
\usepackage[displaymath]{lineno}
\usepackage{float}
\usepackage{booktabs}
\usepackage{bm}
\usepackage{multirow}
\usepackage{makecell}
\usepackage[colorlinks,
linkcolor=red,
anchorcolor=red,
citecolor=red
]{hyperref}

\numberwithin{equation}{section}

\def\b{\boldsymbol}

\def\e{\epsilon}

\def\Z{\mathbb{Z}}
\def\R{\mathbb{R}}
\def\E{\mathbb{E}}

\def\cO{\mathcal{O}}

\DeclareMathOperator{\erf}{erf}
\DeclareMathOperator{\erfc}{erfc}

\newcommand{\tcb}[1]{\textcolor{blue}{#1}}

\newcommand*\samethanks[1][\value{footnote}]{\footnotemark[#1]}

\newtheorem{remark}{Remark}
\newtheorem{proposition}{Proposition}

\begin{document}

	\title{A random batch Ewald method for charged particles in the isothermal-isobaric ensemble}
	
	\author[1]{Jiuyang Liang}
	\author[3]{Pan Tan}
	\author[2,3]{Liang Hong}
	\author[1,2]{Shi Jin}
	\author[1,2]{Zhenli Xu\thanks{Corresponding authors\\
	Email addresses: xuzl@sjtu.edu.cn, leili2010@sjtu.edu.cn}}
	\author[1,2]{Lei Li\samethanks[1]}
	\affil[1]{School of Mathematical Sciences, Shanghai Jiao Tong University, Shanghai, 200240, P. R. China}
	\affil[2]{Institute of Natural Sciences and MOE-LSC, Shanghai Jiao Tong University, Shanghai, 200240, P. R. China}
	\affil[3]{School of Physics and Astronomy, Shanghai Jiao Tong University, Shanghai 200240, P. R. China}

	\date{}
	\maketitle
	
We develop an accurate, highly efficient and scalable random batch Ewald (RBE) method to conduct simulations in the isothermal-isobaric ensemble (the NPT ensemble) for charged particles in a periodic box. 
After discretizing the Langevin equations of motion derived using suitable Lagrangians, the RBE method builds the mini-batch strategy into the Fourier space in the Ewald summation for the pressure and forces so the computational cost is reduced from $\mathcal{O}(N^2)$ to $\mathcal{O}(N)$ per time step.
We implement the method in the LAMMPS package and report accurate simulation results for both dynamical quantities and statistics for equilibrium for typical systems including all-atom bulk water and a semi-isotropic membrane system. Numerical simulations on massive supercomputing cluster are also performed to show promising CPU efficiency of RBE.
	
\section{Introduction}
	
	Molecular dynamics (MD) simulation has been a powerful tool for studying a broad range of systems in physical, chemical, biological, pharmaceutical and materials sciences at the nano/micro scale \cite{brunger1987crystallographic,karplus1990molecular,hollingsworth2018molecular,yamakov2002dislocation,allen2017computer}.
	The common purpose of such a simulation is the configuration sampling by solving the equations of motion, 	and the calculation of equilibrium and dynamical quantities from the ensemble averages.  The equations of motion are coupled to external baths, modelled by thermostats and/or barostats. In this paper, we consider the isothermal-isobaric ensemble (or NPT ensemble) , where the temperature and pressure in the bath are constant. 
The NPT requires to retain constant particle number, external pressure and temperature in bath during the simulation because many experiments in laboratory are carried out with the same conditions. Hence, this ensemble is frequently required in simulating solvated proteins, membranes and virus \cite{tobias1997atomic}. 

The thermostats and barostats are used for modelling constant external pressure and temperature in simulations. Various thermostats and barostats have been proposed, though not all of them can generate the correct distribution for the NPT ensemble. Pressure control with barostats was first introduced in the work of Andersen \cite{andersen1980molecular}, where the (internal) pressure is adjusted by allowing the volume of the simulation cell to fluctuate according to the difference to the external pressure. The temperature 
control is achieved by a random velocity resampling to model the collision with the bath. Andersen's method cannot give the dynamical information like mean square displacement (MSD).
 Extension of the pressure control to periodic cells of arbitrary shape was then studied by Parrinello and Rahman  for the NHP ensemble (i.e. constant particle number, external pressure and enthalpy) \cite{parrinello1980crystal,parrinello1981polymorphic}. 
Alternatively, the temperature and pressure control can be conducted by using the method of Berendsen \cite{berendsen1984molecular}, in which the velocities of the particles and the volume of the box are altered in rates proportional to the difference between the instantaneous quantities (temperature and pressure) and the target ones. However, the Berendsen's method cannot generate the correct distribution function for the NPT ensemble. Later,  a stochastic version of  the Berendsen thermostat was proposed in \cite{bussi2007canonical,bussi2009isothermal}. 	Similar idea for the barostats by random fluctuation of volume was conducted in the stochastic cell rescaling method \cite{bernetti2020pressure}.	The accuracy of the stochastic cell rescaling method has been well demonstrated in many problems, though 	no theoretical demonstration has been given yet.

The Nos\'e-Hoover thermostat \cite{hoover1985canonical,nose1984molecular,nose1983constant,nose1984unified}
	was introduced to replace the additional stochastic collisions used for the temperature control in Andersen's work.	In this type of thermostat, a new auxiliary variable is introduced to model the effects of bath. The Nos\'e-Hoover thermostat has better performance in preserving dynamical properties since the perturbation to the system is made through a single auxiliary degree of freedom, but it is inefficient for equilibration due to the lack of ergodicity \cite{frenkel2001understanding}.
	This can be partially solved by using the Nos\'e-Hoover chains \cite{martyna1992nose} approach with additional number of parameters. Combination of the Nos\'e-Hoover thermostat together with its extensions and various barostats leads to a variety of different methods \cite{melchionna1993hoover,martyna1994constant,martyna1996explicit,
	sturgeon2000symplectic,kalibaeva2003constant,tuckerman2006liouville}. Among them, the MTK algorithm \cite{martyna1994constant,martyna1996explicit} and its variants have been widely used for the NPT ensemble.

The Langevin dynamics \cite{frenkel2001understanding} is another frequently used thermostat, where the dissipative forces and the noise are added to the Hamiltonian dynamics, with the fluctuation-dissipation relation satisfied, for the control of the temperature. The dynamics satisfies the ergodicity so that the correct temperature and fluctuations can be achieved. 
The Langevin piston scheme and the improvements \cite{feller1995constant,bussi2007accurate,kolb1999optimized,gronbech2014constant,di2015stochastic} use the Langevin thermostat and certain barostats to achieve the temperature and pressure control. Such methodology has later been extended to fully flexible cell motions \cite{quigley2004langevin,quigley2005constant,gao2016sampling}.

In practical simulations, it is of great importance that the instantaneous pressure of the system can be accurately and efficiently calculated. The development of accurate, convenient and general formulations of pressure and stress tensor is nontrivial. The instantaneous internal pressure of the system for nonperiodic cells can be computed using the virial theorem, where one uses pairwise interaction \cite{frenkel2001understanding,allen2017computer}. 
When it comes with the periodic cells, the efficient computation of the correct virial is tricky. For short-range interactions, 
one may consider the nearest image and obtain a formula similar to the one for nonperiodic cells. Bekker {\it et al.} \cite{bekker1993gromacs} shows that	it is possible to extract the virial calculation from the inner loop to compute it efficiently.
For general interactions, there is some extra correction term for the pressure tensor under periodic boundary conditions, as derived in \cite{louwerse2006calculation,thompson2009general}. Using the formulas in \cite{louwerse2006calculation,thompson2009general} to compute the pressure for long range interactions like the Coulomb interaction is expensive as the computational cost is of $\mathcal{O}(N^2)$ where $N$ is the number of particles.
 For Coulomb interaction, a possible way to resolve this is the classical Ewald summation method \cite{ewald1921berechnung}, in which the long range part can be computed in Fourier space and the resulted pressure formula can be computed in less effort \cite{brown1995general}. Such virial formulas can be computed with $\mathcal{O}(N\log N)$ complexity via some  lattice-based technique using the fast Fourier transform (FFT)  \cite{sega2016pressure}.
However,  costly communication-intensive tasks will significantly reduce the scalability in parallel computation using such methods.   An efficient expression of the virial calculation together with a fast and scalable Coulomb solver remains a significant topic as this calculation tends to dominate the total runtime of MD simulations.
	
	In this paper, we propose a random batch Ewald (RBE) algorithm to conduct accurate and efficient simulations in the NPT ensemble for charged systems, which also exhibits good scalability in parallel computation.
We use the Langevin dynamics for the thermostat and the equations of motion together with the barostat are derived
from some suitable Lagrangians. The derivation of the equations is performed for general cases of the equations of motion, including isotropic, semi-iostropic, anisotropic, and fully flexible cell fluctuations. The connection of the Langevin dyanmics derived here to the one in the cell rescaling method \cite{bernetti2020pressure} is clarified.
The instantaneous pressure for the Coulomb interaction will be computed using the Ewald summation. To overcome the 
scalability issues for the mesh-based methods, we adopt the idea in recent-developed RBE summation \cite{jin2021random,liang2021Sta}  for canonical ensemble. The RBE method is based on the Ewald summation, but avoids the use of the FFT. Instead, it applies a random mini-batch importance sampling strategy on the Fourier space to approximate the force and pressure contribution from the long-range part. Similar as the RBE method for canonical ensemble, in the RBE method developed for NPT ensemble, the complexity for computing pressure is reduced to $\mathcal{O}(N)$ per time step, and the communication cost between cores is also significantly reduced. The method can be viewed as certain Monte Carlo type methods, and the importance sampling strategy makes the variance insensitive to particle number $N$ and thus is accurate and efficient, shown in numerical simulations. 

This RBE algorithm for the NPT ensemble is implemented and systematically tested 
	in a modified version of the MD package LAMMPS \cite{plimpton1995fast}, and is demonstrated to produce rapid convergence and correct physical quantities. Simulations are conducted on bulk water system 
	with isotropic fluctuation, revealing that the spatiotemporal information on all time and length scales and 
	the thermodynamical quantities are quantitatively reproduced. Moreover, the semi-isotropic version of the RBE is tested 
	on a membrane simulation, which illustrates the high accuracy and stability of the method. Simulations on massive supercomputing cluster show promising CPU efficiency and scalability of this RBE algorithm.
	
	The rest of this paper is organized as follows. The distribution functions for frequently-used types of cell's fluctuations 
	under NPT ensemble are discussed in Section \ref{distributionfunction}. The development of
	NPT Langevin equations of motion for cubic cell and the measurement of the pressure are discribed in Section \ref{sec:langfixedshape}. In Section \ref{sec:randombatch}, the random batch Ewald method is presented to approximate the long-range interaction and instantaneous pressure  together with the complexity and rationalization analysis. Section \ref{sec:numerics}
	contains simulation results of our new scheme on all-atom bulk water and membrane systems. Conclusions are made in Section \ref{conclusion}.  Extensions of equations of motion to general cells under Langevin dynamics, discussions on the Hoover-type equations of motion, and the RBE method for general cells are given in the Appendix section.
	
	\section{Distribution functions under NPT ensemble}\label{distributionfunction}
	
	For a system in contact with a heat bath, the probability of a microstate of the system under the NPT ensemble that achieves
	energy $E$ and volume $V$ is given by (see, e.g., \cite[Appendix H, Eq. (18)]{beale2011statistical} or \cite[Sec. 8.8]{leimkuhler2016molecular})
	\begin{gather}\label{eq:probstate}
		\mathrm{Prob}\propto \exp(-\beta(E+PV)),
	\end{gather}
	where $P$ is the external pressure exerted on the system from the heat bath and $\beta=(k_{\text{B}}T)^{-1}$ with $k_{\text{B}}$ the Boltzmann constant and $T$ the temperature is the inverse thermal energy. Such an ensemble in which the probability for a microstate is given by \eqref{eq:probstate} is called the isothermal-isobaric ensemble, or the NPT ensemble as the partition function is a function of the particle number $N$, the external pressure $P$ and the temperature $T$.

	We consider a cell tensor
	\begin{gather}
		\bm{h}=[\bm{h}_1, \bm{h}_2, \bm{h}_3]\in \R^{3\times3},
	\end{gather}
	which describes both the shape and the size of the cell. Write $\bm{h}=V^{1/3}\bm{h}_0$ where the determinant $\mathrm{det}(\bm{h}_0)=1$ so that $\bm{h}_0$ gives the shape of the cell and $V$ is the volume. Together with constant particle number $N$ in the system, the Cartesian momenta and coordinates of the $N$ particles are denoted by $\left\{\b{p}_{1},\cdots,\b{p}_{N}\right\}\equiv \b{p}$, and $\left\{\b{r}_i,\cdots,\b{r}_N\right\}\equiv \b{r}$, respectively. It will also be convenient to write $\bm{r}_i\in\R^3$ as
	\begin{gather}
		\bm{r}_i=\sum_{j=1}^{3} s_i^{(j)}\bm{h}_{j}, \bm{s}_i\in \R^3,
	\end{gather}
	so that $\bm{s}_i$ is the reduced position vector and $s_i^{(j)}$ is its $j$th component.

To find the formula for the partition function or the distribution function in terms of some macroscopic variables, we need to find the density of states. The probability that the system is found at a particular macrostate with parameters $(\bm{r}, \bm{p}, V, \bm{h}_0)$ is obtained by multiplying the probability \eqref{eq:probstate} with the density of states. We denote that $d\b{p}=d\b{p}_1\cdots d\b{p}_N$ and $d\b{r}=d\b{r}_1\cdots d\b{r}_N$. By the classical quasi-classical result, the number of states in
	the volume element $d\b{r}d\b{p}$ is given by $d\b{r}d\b{p}/(2\pi\hbar)^{3N}$ \cite{landau2013course},
	where $N$ is the number of particles, which implies that the microstates are uniformly distributed in the phase space described by $\{\bm{r}, \bm{p}\}$.
	Consider a fixed shape $\b{h}_0$. Next, one needs to count the number of states in $(V, V+dV)$. As mentioned in the footnotes of \cite[Section 5.4.1]{frenkel2001understanding}, this is tricky as the volume cannot be counted directly \footnote{By ``cannot be counted'', it means that the volume is not discretized in the quantum regime so one cannot estimate the total number of states in $(V, V+dV)$.}. In Ref. \cite{koper1996length}, it has been justified by counting the quantum states that one should use
	$\beta p''(V)dV$ for the integral, where $p''(V)$ is the pressure for our system in thermal equilibrium under canonical ensemble. 
	In most cases, the fluctuation of volume is not big and one may get good results by setting $p''(V)$ be a constant. With this approximation, the probability to find the system at  $(\bm{r}, \bm{p}, V)$ under the NPT ensemble with the fixed shape is thus given by
	\begin{gather}\label{eq:distfixedshape}
		dw \propto \exp(-\beta (E+P V))d\bm{r}d\bm{p}dV .
	\end{gather}
	In other words, the partition function under the NPT ensemble for fixed shape is given by
	\begin{gather}\label{DeltaNPT}
		\Delta(N, P, T)= \int \exp(-\beta (E+P V)) d\bm{r} d\bm{p} dV
		=\int_0^{\infty} dV \exp(-\beta PV) Q(N, V, T),
	\end{gather}
	where $Q(N, V, T)=\int_{D(V)\times \R^{3N} }\exp(-\beta E(\{\bm{r}, \bm{p}\}))d\b{r} d\b{p}$ is the parition function for canonical ensemble, with $D(V)$ being the spatial domain for the system \cite{martyna1994constant}.

	In some applications, the shape of the cell can change, and one may have some parameters to parametrize $\b{h}$. In this case, one should put weights for those $\b{h}_0$. Similarly, for the variable $V$, as commented in the footnotes of \cite[Section 5.4.1]{frenkel2001understanding}, one cannot count $\b{h}_0$. Usually, one may assume the uniform weight for shapes (with $\det(\b{h}_0)=1$) under the natural integral element for the parameters parametrizing $\b{h}_0$.  Here, some frequently-used examples are considered:
	\begin{enumerate}[(i)]
		\item Consider that the cell is a rectangular box, where the bottom is a square with area $A$ and the height is $L$, i.e., the pressure in $x$ and $y$ direction are (isotropically) coupled and the anisotropy is applied in the $z$ direction, with $V=AL$. This case is refered to as the semi-isotropic fluctuation, which is widely used to simulate membrane systems \cite{kandt2007setting,ivanova2018testing}. Denoting that $A_0=V^{-2/3}A$ and $L_0=V^{-1/3}L$, we may put the uniform weight so that the integration measure for the shapes is given by $dA_0dL_0\delta(A_0L_0-1)$ where $\delta(\cdot)$ denotes the one-dimensional Dirac delta function. Then, the partition function is given by
		\begin{gather}
			\Delta(N, P, T) = \int_0^{\infty} dV \exp(-\beta PV)\widetilde{Q}(N, V, T),
		\end{gather}
		where
		$\widetilde{Q}(N, V, T)=\int_{\R_+^2} dA_0dL_0\delta(A_0L_0-1) Q(\b{h})$ and
		\begin{gather}
			Q(\b{h})=\int_{D(\b{h})\times\R^{3N}}\exp(-\beta E)d\b{r} d\b{p},
		\end{gather}
	is	the parition function for canonical ensemble with $D(\b{h})$ being the domain defined by the cell tensor $\bm{h}$.
		Then, one finds
		\begin{gather}
			\Delta(N, P, T)= \int  \exp(-\beta(E+PAL)) d\b{r}d\b{p} dA dL,
		\end{gather}
		so that the distribution function is determined by
		\begin{gather}
			dw \propto \exp(-\beta(E+PAL))d\bm{r}d\bm{p}dAdL.
		\end{gather}
		We remark that $E$ may also depend on $\b{h}$, $A$ and $L$. If the non-zero surface-tension $\gamma_{0}$ exists, i.e., one wants to sample the constant surface-tension ensemble $NP\gamma_{0}T$ \cite{zhang1995computer}, then $E=K+U-\gamma_0 A$, where $K$ and $U$ are the kinetic and potential energies, and thus
		\begin{equation}\label{semiisotropic}
			dw \propto \exp(-\beta(K+U+PAL-\gamma_0 A))d\bm{r}d\bm{p}dAdL.
		\end{equation}
		
		\item If the cell is a rectangular box with three side lengths being $\b{\ell}=\{\ell_{j},~j=1,2,3\}$, referring to the case of anisotropic fluctuation, one may assign the measure similarly as
		$\prod_{j}d\ell_{0,j}\delta(\prod_{j}\ell_{0,j}-1)$ where $\ell_{0,j}=V^{-1/3}\ell_{j}$, and derive that the distribution function using variables $\{\b{r}, \b{p}, \b{\ell}\}$ is determined by
		\begin{gather}\label{eq:distrectangular}
			dw \propto \exp\bigg{(}-\beta \bigg{(}E+P\prod_{j} \ell_{j} \bigg{)}\bigg{)}d\bm{r}d\bm{p}d\b{\ell}.
		\end{gather}
		
		\item Consider the general fully flexible cell $\b{h}$. One may put $d\b{h}_0 \delta(\det(\b{h}_0)-1)$ for the shapes, where $d\b{h}=\prod_{j,\eta=1}^3dh_{j\eta}$. Following Martyna {\it et al.} \cite{martyna1994constant},
		the partition function is formulated as
		\begin{gather}
			\begin{split}
				& \int_{[0,\infty)\times \mathbb{R}^{\tilde{d}^2}} dV d\b{h}_0 \delta(\det(\b{h}_0)-1)\exp(-\beta PV)Q(\b{h}) \\
				&= \int_{[0,\infty)\times \mathbb{R}^{\tilde{d}^2}} dV d\b{h}V^{-\tilde{d}}  V\delta(\det(\b{h})-V)\exp(-\beta PV)Q(\b{h})\\
				&=\int_{\det(\b{h})>0} d\b{h} \det(\b{h})^{1-\tilde{d}}\exp(-\beta P\det(\b{h}))Q(\b{h}),
			\end{split}
		\end{gather}
		where the dimension $\tilde{d}=3$, and $\{\b{h}: \det\b{h}>0\}$
		is a subdomain of $\R^{\tilde{d}^2}$.
	\end{enumerate}
	
	\begin{remark}\label{rmk:extratermfully}
		Let us consider (iii) again.  When we use $d\b{h}=\prod_{j\eta}dh_{j\eta}$, the distribution function becomes \cite{martyna1994constant}
		\begin{equation}\label{flexible}
			\det(\b{h})^{1-\tilde{d}}\exp(-\beta (E+P\det(\b{h}))) d\b{r} d\b{p} d\b{h}.
		\end{equation}
		The extra factor $\det (\b{h})^{1-\tilde{d}}$ will give an extra term in the Hamiltonian constructed for the equations of motions, as will be seen in Section~\ref{extensions}. Such a term is due to the fact that $d\b{h}$ is not the infinitesimal volume for the system.
	\end{remark}
	
	As a comment, the weight for the shape $\b{h}_0$ does not have to be
	$d\b{h}_0 \delta(\det(\b{h}_0)-1)$. In fact, one may integrate $\delta(\det(\b{h}_0)-1)$ against any $\nu$-form $\omega_{\b{h}_0}$.
	In this case,
	\begin{gather}
		\begin{split}
			& \int_{[0,\infty)\times \mathbb{R}^{\nu}}  \delta(\det(\b{h}_0)-1)\exp(-\beta PV)Q(\b{h}) dV \omega_{\b{h}_0} \\
			&= \int_{[0,\infty)\times \mathbb{R}^{\nu}}  V^{-\nu/\tilde{d}}  V\delta(\det(\b{h})-V)\exp(-\beta PV)Q(\b{h}) dV \omega_{\b{h}}\\
			&=\int_{\R^{\nu}\cap \{\det h>0\}} \det(\b{h})^{1-\nu/\tilde{d}}\exp(-\beta P\det(\b{h}))Q(\b{h})  \omega_{\b{h}}.
		\end{split}
	\end{gather}
	For the cubic box, one may take $\nu=\tilde{d}=3$ and $\omega_{\b{h}}=dh_{11}dh_{22}dh_{33}$ so that the distribution function agrees with \eqref{eq:distrectangular}. Similarly, for the cubic box, we may take $\nu=1$ and $\omega_{\b{h}}=dh_{11}$, $h_{11}=L=V^{1/3}$ and the distribution function is determined by
	\begin{gather*}
		dw\propto\exp(-\beta(E+PV)) (L^{2}dL)d\b{r}d\b{p}\propto \exp(-\beta (E+P V))d\b{r}d\b{p} dV.
	\end{gather*}
	This agrees with Eq.~\eqref{eq:distfixedshape}.

	To generate the NPT ensemble, we need the equations of motion under certain thermostats to have the desired distribution functions above.
	There are already many classical approaches \cite{andersen1980molecular,berendsen1984molecular,parrinello1981polymorphic,gao2016sampling,martyna1994constant,bussi2009isothermal,bernetti2020pressure,feller1995constant,hoover1986constant} for generating the equations of motion.
	We choose to derive the equations of motion following the strategies in \cite{parrinello1981polymorphic,gao2016sampling}. Generally speaking, we
	turn the parameters for the cell $\b{h}$ into the dynamical variables by building them into the Lagrangians. From the Lagrangians, one may find the corresponding Hamiltonians, and then generate the desired ensembles using the traditional thermostats. We will mainly focus on the cells with fixed shape $\b{h}_0$ under the Langevin thermostat in Section \ref{sec:langfixedshape} for illustration of the derivation and how the RBE algorithm works in Section \ref{sec:randombatch}.
	We then discuss the extensions in Appendix~\ref{extensions} to derive the equations of motion for other changing shapes and for Hoover-type thermostats in Appendix~\ref{app:hoover}. The discussion on RBE method for general cases with changing shapes is given in Appendix \ref{app:rbegeneral}.

	\section{The Langevin equations of motion}\label{sec:langfixedshape}
	
	As mentioned above, we consider the cells with fixed shapes (i.e., $\b{h}_0$) under the Langevin dynamics \cite{frenkel2001understanding}. The extensions to other cases can be found in Appendix \ref{extensions}.  To start with, we first consider the Lagrangians corresponding to the NHP ensemble ($H$ refers to the enthalpy) \cite{parrinello1981polymorphic,nose1983constant}, and then derive the equations of motion under the Langevin dynamics using the first principle.
	
	\subsection{The Hamiltonian and the microscopic pressure}

	Consider the reduced position vector
	$\b{s}_i=V^{-1/3}\b{r}_i$, which gives the relative position in the cell. We use the variables $\b{s}_i$ and $V$ to construct the Lagrangian. Motivated by the discussion in \cite{parrinello1981polymorphic} and \cite[Section 2.3]{nose1983constant}, we consider the Lagrangian of the following form
	\begin{gather}\label{eq:Lagrangianforcubic}
		\mathscr{L}(\{\b{s}_i, \dot{\b{s}}_i\}, V, \dot{V})
		=\sum_i \frac{1}{2}m_i V^{2/3}|\dot{\b{s}}_i|^2
		+\frac{1}{2}M(\dot{V})^2 -(U(\{V^{1/3}\b{s}_i\}; V) + PV),
	\end{gather}
	where $m_i$ is the mass of the $i$th particle, $M$ is the virtual mass of the cell, and $U(\{V^{1/3}\b{s}_i\}; V)$ is the potential energy of the system. Here, we assume that $U(\{V^{1/3}\b{s}_i\}; V)$ not only depends on the positions of the particles but also on the volume of the cubic box $V$. This will be clear when we discuss the periodic boundary conditions later.

	We note that in Eq.\eqref{eq:Lagrangianforcubic}, the variable $V^{1/3}\dot{\b{s}}_i=\b{h}\dot{\b{s}}_i$ is the artificial velocity used to generate
	the NHP ensemble for isotropic barostat \cite{parrinello1981polymorphic}. The ``kinetic energy'' for the $i$th particle is chosen to be $m_{i}(\b{h}\dot{\b{s}}_{i})^2/2$ instead of $m_{i}[(d/dt)(\b{h}\b{s}_i)]^2/2$, as has been verified already in \cite{parrinello1981polymorphic,gao2016sampling}, to yield the correct isobaric ensembles. In fact, though $\b{h}\dot{\b{s}}_i$ is not the physical velocity, the equations derived from this Lagrangian can be converted back to equations with $\b{r}_i=\b{h}\b{s}_i$ and $\b{p}_i=m_i \b{h}\dot{\b{s}}_i$. These equations for $\b{r}_i$ can be viewed as the physical equations and the ensemble generated is the correct isobaric ensembles. Besides, if $\b{h}$ is varying in a slower scale compared to the motion of particles, $\b{h}\dot{\b{s}}_i$ is also approximately the physical velocity.
	
	\begin{remark}
		As a side remark, the kinetic energy used for the cell $\b{h}$ is different from the one used for the fully flexible one, where one uses
		$(1/2)M\dot{\b{h}}:\dot{\b{h}}$. If we use an analogue here,  the kinetic part for the cell will be
		$(1/2)M (\dot{L})^2\sim V^{-4/3}\dot{V}^2$ with $L=V^{1/3}$. This will give complicated equations of motion. As commented in \cite{nose1983constant}, the equations of motion derived from the Lagrangian in Eq.~\eqref{eq:Lagrangianforcubic} will be cleaner.
	\end{remark}

	Define the conjugate variables for $\b{s}_i$ and $V$,
	\begin{gather}
		\b{p}^s_i=V^{2/3}m_i \dot{\b{s}}_i,~~~p^V=M\dot{V},
	\end{gather}
	so that the ``Hamiltonian'' is given by
	\begin{gather}
		\mathscr{H}(\{\b{s}_i, \b{p}^s_i\}, V, p^V)
		=\sum_i \frac{V^{-2/3}|\b{p}_i^s|^2}{2m_i}
		+\frac{(p^V)^2}{2M}+(U(\{V^{1/3}\b{s}_i\}; V) + PV).
	\end{gather}
	We can see that the Hamiltonian here is simply $E+PV+(p^V)^2/(2M)$, i.e., the enthalpy plus $(p^V)^2/(2M)$. Hence, the Hamilton ODEs under this Hamiltonian will generate an approximate NHP ensemble
	as in the original work \cite{parrinello1981polymorphic}, as $(p^V)^2/(2M)$ is small compared to the kinetic energy of the particles.
	Considering the Boltzmann distribution $\exp(-\beta\mathscr{H})$, the marginal distribution for $\b{r}_i, \b{p}_i$ and $V$ is exactly the NPT distribution above since the Jacobian for the transform $(\b{s}_i, \b{p}_i^s)\to (\b{r}_i, \b{p}_i)$ is $1$. Hence, in principle, any thermostat that can generate the distribution
	$\exp(-\beta\mathscr{H})d\b{p}_i d\b{r}_i dV$ can give the correct NPT ensemble.
	
	According to standard statistical physics, the pressure of the system is given by
	\begin{gather}\label{eq:pressure1}
		\widetilde{P}=\frac{1}{\beta}\left(\frac{\partial \log Q}{\partial V}\right)\bigg{|}_T,
	\end{gather}
	where one recalls
	\begin{gather}
		Q(N, V, T)=\int \exp\left(-\beta\left(\sum_i \frac{|\b{p}_i|^2}{2m_i}+U(\{\b{r}_i\}; V)\right)\right)d\b{r}_i d\b{p}_i.
	\end{gather}
	The bound for $\b{r}_i$ depends on $V$, so the derivative can be taken more easily by changing the variables to $\b{s}_i$ and $\b{p}_i^s$.
	Denoting $K(V, \b{p}_i^s):= V^{-2/3}\sum_i|\b{p}^s_i|^2/(2m_i)$ be the kinetic energy of the system, one then has
	\begin{gather}
		\widetilde{P}=-\frac{1}{Q(N, V, T)}\int \exp\left[-\beta\left(\sum_i \frac{V^{-2/3}|\b{p}_i^s|^2}{2m_i}
		+U(\{V^{1/3}\b{s}_i\}; V)\right)\right]\frac{\partial (K+U)}{\partial V}d\b{s}_i d\b{p}_i^s.
	\end{gather}
	From here, we infer that the instantaneous pressure can be written as
	\begin{gather}\label{eq:inspressure2}
		P_{\mathrm{ins}}=\frac{\partial(-K-U)}{\partial V}\bigg{|}_{\b{s}_i\leftarrow V^{-1/3}\b{r}_i}=\frac{1}{3V}\left[\sum_i \frac{|\b{p}_i|^2}{m_i}-3V\frac{\partial U(\{V^{1/3}\b{s}_i\}, V)}{\partial V}\bigg{|}_{\b{s}_i\leftarrow V^{-1/3}\b{r}_i}\right],
	\end{gather}
	where the partial derivative now is taken with variable $\b{s}_i$ fixed.

	By computing the derivative more explicitly, one finds that the instantanenous pressure Eq.\eqref{eq:inspressure2} is also given by
	\begin{gather}\label{eq:inspressure1}
		P_{\mathrm{ins}}:=\frac{1}{3V}\sum_i \frac{|\b{p}_i|^2}{m_i}
		-\frac{1}{3V}\b{r}_i\cdot \nabla_{\b{r}_i}U(\{\b{r}_i\}; V)-\frac{\partial }{\partial V}U(\{\b{r}_i\}; V).
	\end{gather}

	\begin{remark}\label{rmk:extravirial}
		In the pressure expression here, one finds that the virial in the periodic setting has an extra term corresponding to $\partial U/\partial V$. This is due to the extra momentum flux by the image motion caused by this cell reshaping. Such kind of momentum flux is neither due to the forces exerted on the particles nor due to the motion of the particle itself. In fact, such a correction due to periodic boundary conditions has been investigated in literature \cite{thompson2009general}.
	\end{remark}

	We note that the explicit method for evaluating the correction $(\partial/\partial V)U(\{\b{r}_i\}; V)$ discussed in \cite{thompson2009general} is not very convenient for our setting as we will use the Ewald summation \cite{ewald1921berechnung} to get the expression for $U$ and compute the derivatives following \cite{brown1995general}.  Hence, for our purpose,
	expression Eq.\eqref{eq:inspressure2} is more useful than Eq.\eqref{eq:inspressure1} in the Ewald summation discussed in Section \ref{Ewaldsum}.

	Using the expression \eqref{eq:pressure1}, one may verify the two pressure virial theorems:
	\begin{gather}
		\begin{split}
			& \left< P_{\mathrm{ins}}\right>=P,\\
			&	\left< P_{\mathrm{ins}}V\right>=P\left<V\right>-k_{\mathrm{B}}T.
		\end{split}
	\end{gather}
	where $\left<\cdot\right>$ denotes the NPT ensemble average. Note that the first theorem relates the internal and external pressures and the second theorem could give the equation of states. The justification can be done by following the appendix B of \cite{martyna1994constant}.

	At last, one has by definition that
	\begin{gather}
		\frac{\partial \mathscr{H}}{\partial V}=-P_{\mathrm{ins}}+P,
	\end{gather}
	which will be used in the equations of motions for $p^V$.
	
	\subsection{Equations of motion for isotropic case}\label{equationofmotion}
	
	Considering the Langevin dynamics with friction and noise terms added into the Hamilton ODEs which satisfy the fluctuation-dissipation relations
	so that the equilibrium distribution is the one we desire. In particular, suppose that $p$ and $q$
	are conjugate variables in the Hamiltonian, then one may construct equations as
	\begin{gather}\label{eq:Langevingeneral}
		\begin{split}
			& d q=\frac{\partial \mathscr{H}}{\partial p}\,dt,\\
			& d p=\left(-\frac{\partial \mathscr{H}}{\partial q}-\Gamma(q) \frac{\partial \mathscr{H}}{\partial p}\right)\,dt+\Sigma(q)\, d W_p,
		\end{split}
	\end{gather}
	where $\Sigma(q)\Sigma^T(q)=2k_{\text{B}}T \Gamma(q)$ is the fluctuation-dissipation relations and $W_{p}$ denotes the standard Wiener processes.
	Note that we allow $\Sigma$ to depend on $q$. It can be verified easily using the Fokker-Planck equation that the invariant measure is $\propto \exp(-\mathscr{H}/(k_{\text{B}}T))dpdq$ \cite{weinan2019applied}.

	Using the general Langevin equations Eq.~\eqref{eq:Langevingeneral}, one may construct the equations of motion for the cubic boxes:
	\begin{gather}\label{LangevinEquation}
		\begin{split}
			& d\b{s}_i=V^{-2/3}\frac{\b{p}^s_i}{m_i}\,dt,\quad d\b{p}_{i}^s=\left( -V^{1/3}\nabla_{\b{r}_i}U
			-\Gamma_i V^{-2/3}\frac{\b{p}^s_i}{m_i}\right)\,dt+\Sigma_i\, d\b{W}_i,\\
			& dV=\frac{p^V}{M}\,dt,
			\quad d p^V=\left(P_{\mathrm{ins}}-P-\tilde{\gamma}\frac{p^V}{M}\right)\,dt +\hat{\sigma}\, d W_{V}.
		\end{split}
	\end{gather}
	Note that $\b{W}_i$ and $W_{V}$ in Eq.\eqref{LangevinEquation} are independent for different particles and the cell variable $V$. We set $\Sigma_i=V^{1/3}\sqrt{2k_BT\gamma_i m_i}$ and
	$\Gamma_i=(2k_{\text{B}}T)^{-1}\Sigma_i\Sigma_i^T=V^{2/3}\gamma_i m_i$, where the friction coefficient $\gamma_i$ is set to be constant for convenience.
	Choosing such $\Sigma_i$ can make the equations for the variables $\b{r}_i, \b{p}_i$ clean as we shall see below.
	We also choose $\hat{\sigma}=\sqrt{2\gamma(V)k_BT M}$
	so that $\tilde{\gamma}=\gamma(V)M$.
	The coefficient $\gamma(V)$ is allowed to depend on $V$ so that this artificial
	friction coefficient may yield better ability to adjust $V$ under the isobaric ensemble, as we will discuss below.

	Changing back to the original variables $\b{r}_i=V^{1/3}\b{s}_i$ and $\b{p}_i=V^{-1/3}\b{p}^s_i$, one has
	\begin{gather}\label{24}
		\begin{split}
			& \dot{\b{r}}_i=\frac{\b{p}_i}{m_i}+\frac{\dot{V}}{3V} \b{r}_i,~~~\dot{\b{p}}_{i}= -\nabla_{\b{r}_i}U-\frac{\dot{V}}{3V}\b{p}_i
			-\gamma_i \b{p}_i+\sqrt{2k_{\mathrm{B}}Tm_i\gamma_i} \dot{\b{W}}_i,\\
			& \dot{V}=\frac{p^V}{M},
			~~~\dot{p}^V=P_{\mathrm{ins}}-P-\gamma(V)p^V +\sqrt{2k_BT\gamma(V) M} \dot{W}_{V}.
		\end{split}
	\end{gather}
	The equations for $\b{r}_i$ and $\b{p}_i$ have clear physical interpretations compared to the equations for $s_i$ and $p_i^s$.
	As can be seen, in the equation of $\b{r}_i$, besides $\b{p}_i/m_i$, one has the extra term associated with $\dot{V}/(3V)$, which is known as the compressibility.
	It means that the position of the particle is also scaled with the cell, and this coincides with the equations of motion in the Berendesen method \cite{berendsen1984molecular} and the Langevin piston method \cite{feller1995constant}.
	These equations for variables $\b{r}_i, \b{p}_i$ clearly have their physical interpretations. Hence, even though we used the aritificial kinetic energy in the Lagrangian, these equations of motion generate the correct NPT ensemble, and $\b{p}_i$
	can be regarded as physical momentum, forgetting the variables $\b{s}_i$ and $\b{p}_i^s$.

	Now, we take a look at the connection between the equations above and some current approaches for pressure controlling. We note that the differential equation of the cell rescaling (Eq.~(5) in Ref.~\cite{bernetti2020pressure}) is an asymptotic case of the Langevin equations (for the cubic cells) Eq.~\eqref{24}. Following the Smoluchowski-Kramers approximation given in Ref.~\cite[Theorem~1]{hottovy2015smoluchowski}, under the assumptions that the instantaneous pressure is a continuously differentiable function and the friction coefficient $\gamma(V)$ is a scalar function of $V$, the cell momentum $p^{V}$ in Eq.~\eqref{24} can be formally eliminated as a fast variable by taking the zero-mass/high-friction limit.
	In fact, setting
	\[
	\tilde{\gamma}(V)=\gamma(V)M,
	\]
	and fixing $\tilde{\gamma}(V)$ as $M\to 0$, one then obtains the well-known over-damped Langevin dynamics
	\begin{equation}
		dV=\left[\dfrac{P_{\text{ins}}-P}{\tilde{\gamma}(V)}-\dfrac{k_{\text{B}}T}{\tilde{\gamma}(V)^2}\dfrac{d\tilde{\gamma}(V)}{dV}\right]dt+\sqrt{\dfrac{2k_{\text{B}}T}{\tilde{\gamma}(V)}}d{W}_{V}.
	\end{equation}
	Here, we may see why varying $\tilde{\gamma}$ may be beneficial in some cases.
	In fact, if $\tilde{\gamma}$ is a constant, the equation is simply
	\begin{gather}
		dV=\gamma_1\left[P_{\text{ins}}-P\right]dt+\sqrt{2k_{\text{B}}T\gamma_1}d{W}_{V},
	\end{gather}
	where $\gamma_1=1/\tilde{\gamma}$. However, if we take $\varepsilon=\log(V)$ and $\tilde{\gamma}(V)=\dfrac{\tau_{P}}{\beta_{T}V}$, where $\beta_{T}$ is an estimate
	of the isothermal compressibility of the system and $\tau_{P}$ is a
	characteristic time associated to the barostat, one obtains
	\begin{equation}\label{overdamped}
		d\varepsilon=-\dfrac{\beta_{T}}{\tau_{P}}\left(P-P_{\text{ins}}\right)dt+\sqrt{\dfrac{2k_{\text{B}}T\beta_{T}}{V\tau_{P}}}dW_{V}.
	\end{equation}
	Eq.~\eqref{overdamped} is exactly Eq.~(5) in Ref.~\cite{bernetti2020pressure}.
	This equation can always guarantee the positivity of $V$.
	
	Referring to Sec.~I of the Supplementary Material in Ref.~\cite{bernetti2020pressure}, the Langevin equations shown in Eq.~\eqref{24} also share the same limiting case of other Langevin piston thermostats \cite{feller1995constant,quigley2004langevin,bussi2009isothermal} for high friction with proper choice of $\gamma(V)$.
	
	\subsection{Discretization of the equations of motion}\label{discretize}
	The evolution of a system governed by the Langevin dynamics Eq.\eqref{24} is equivalently described by the following
	Fokker-Planck equation
	\begin{equation}
		\dfrac{\partial\rho}{\partial t}=\mathscr{F}\rho
	\end{equation}
	where $\rho(t,{\b{r}_i,\b{p}_i}, \b{h}, \b{p}^h)$ is the time dependent probability density defined on the phase space. $\mathscr{F}$ is the infinitesimal generator, which can be factorized as
	\begin{equation}
		\mathscr{F}=\mathscr{F}_R+\mathscr{F}_R^p+\mathscr{F}_V+\mathscr{F}_V^p+\mathscr{F}_O+\mathscr{F}_O^p,
	\end{equation}
	where
	\begin{equation}
		\begin{split}
			\mathscr{F}_R&=\sum_{i}\left[\frac{\b{p}_i}{m_i}+\frac{\dot{V}}{3V} \b{r}_i\right]\cdot\dfrac{\partial}{\partial \b{r}_i},\\
			\mathscr{F}_V&=\frac{p^V}{M}\dfrac{\partial}{\partial V},\\
			\mathscr{F}_O&=\sum_{i}\left[3\gamma_i+\gamma_i\b{p}_i\dfrac{\partial}{\partial\b{p}_i}+k_{\text{B}}Tm_i\gamma_i\dfrac{\partial^2}{\partial\b{p}_i^2}\right],\\
			\mathscr{F}_R^p&=\sum_{i}\left[-\nabla_{\b{r}_i}U-\frac{\dot{V}}{3V}\b{p}_i\right]\cdot\dfrac{\partial}{\partial \b{p}_i},\\
			\mathscr{F}_V^p&=(P_{\mathrm{ins}}-P)\dfrac{\partial}{\partial p^V},\\
			\mathscr{F}_O^p&=\gamma(V)+\gamma(V)p^{V}\dfrac{\partial}{\partial p^{V}}+k_{\text{B}}T\gamma(V)M\dfrac{\partial^2}{\partial(p^V)^2}.
		\end{split}
	\end{equation}
	
	Given this factorization, the single-step propagator $e^{\Delta t\mathscr{F}}$ ($\Delta t$ being the time step) can be split by the Trotter method
	\begin{equation}
		e^{\Delta t\mathscr{F}}=e^{\frac{\Delta t}{2}\mathscr{F}_V^p}e^{\frac{\Delta t}{2}\mathscr{F}_R^p}e^{\frac{\Delta t}{2}\mathscr{F}_V}e^{\frac{\Delta t}{2}\mathscr{F}_R}e^{\Delta t\mathscr{F}_O^p}e^{\Delta t\mathscr{F}_O}e^{\frac{\Delta t}{2}\mathscr{F}_R}e^{\frac{\Delta t}{2}\mathscr{F}_V}e^{\frac{\Delta t}{2}\mathscr{F}_R^p}e^{\frac{\Delta t}{2}\mathscr{F}_V^p}+\mathcal{O}(\Delta t^3),
	\end{equation}
	and this kind of splitting is actually the ``BAOAB'' scheme
	\cite{leimkuhler2013robust}. It was argued that the BAOAB splitting is more accurate than other schemes in the sense of configurational sampling. The action of propagators $e^{\Delta t\mathscr{F}_R}$ and $e^{\Delta t\mathscr{F}_R}$ corresponds to evolve $\b{r}_i$ and $\b{p}_i$ by $\Delta t$ under the ordinary differential equations $\dot{\b{r}}_i=\frac{\b{p}_i}{m_i}+\frac{\dot{V}}{3V} \b{r}_i$ and $\dot{\b{p}}_i=-\nabla_{\b{r}_i}U-\frac{\dot{V}}{3V}\b{p}_i$, respectively. The explicit solution of this kind of ODEs can be found in Appendix B of \cite{gao2016sampling}. The action of propagators $e^{\Delta t\mathscr{F}_O}$ and $e^{\Delta t\mathscr{F}_O^p}$ corresponds to evolving variables $\b{p}_i$ and $p^V$ by $\Delta t$ under the Ornstein-Uhlenbeck process. In general, an Ornstein-Uhlenbeck process $d p=-\gamma p \,d t+\sigma \sqrt{m}\, d w_{t}$ can be explicitly solved by
	\begin{equation}
		p(t)=e^{-\gamma t} p(0)+\frac{\sigma}{\sqrt{2 \gamma}} \sqrt{1-e^{-2 \gamma t}} \sqrt{m} \tilde{R},
	\end{equation}
	with $\tilde{R}$ a random number subject to the normal distribution with vanishing mean and unit variance. Similar discretizations can be derived analogically for other cases. We note that the computations of the force for each particle and the instantaneous pressure should be finished after the second update of $e^{\frac{\Delta t}{2}\mathscr{F}_V}$ and before the second update of $e^{\frac{\Delta t}{2}\mathscr{F}_R^p}$ during each time step.

	\section{The random batch Ewald method for NPT ensemble}\label{sec:randombatch}

	In this section, we introduce the RBE idea and apply it to the Langevin equations we derived above for the case with fixed shape, resulting in an efficient method for MD simulations in NPT ensemble. The discussion on the RBE for general cases can be found in Appendix \ref{app:rbegeneral}.

	\subsection{The Ewald sum for forces and pressure}\label{Ewaldsum}
	
	We first review the classical Ewald summation method \cite{ewald1921berechnung}.  Consider $N$ charged particles (numerical particles for simulation or physical particles) with positions $\bm{r}_i$ and charge $q_i$ ($1\le i\le N$) located in a cubic box with side length $L=V^{1/3}$ and satisfying the electroneutrality condition.
	The interaction between them is given by the Coulomb interaction so that the total potential energy is given by
	\begin{gather}\label{eq:energy}
		U=\frac{1}{2}\sum_{\bm{n}}{}'\sum_{i,j=1}^N \frac{q_iq_j}{|\bm{r}_{ij}+\b{h}\bm{n}|},
	\end{gather}
	where $\b{h}=V^{1/3}I_3$ with $I_3$ the three-dimensional identity matrix and $\bm{r}_{ij}:=\bm{r}_j-\bm{r}_i=\b{h}(\b{s}_i-\b{s}_j)$, $\bm{n}\in \mathbb{Z}^3$ ranges over the three-dimensional integer column vectors and the prime indicates that the case $\bm{n}=\b{0}$ (the zero vector in $\mathbb{Z}^3$) is not included when $i=j$. The Coulomb interaction is known to be of long range due to the decay $1/r$. Thanks to the electroneutrality condition,
	the above series converges conditionally.
	Clearly, if we consider the periodic boundary conditions, the potential not only depends on $\b{r}_i$ but also on $V$ so that
	\begin{equation}
		U=U(\{V^{1/3}\b{s}_i\}; V).
	\end{equation}
	This is different from the cases discussed in \cite{parrinello1981polymorphic,gao2016sampling}.

	In the Ewald summation, the Coulomb kernel can be written as
	\begin{equation}
		\frac{1}{r}=\frac{\erf(\sqrt{\alpha}r)}{r}+\frac{\erfc(\sqrt{\alpha}r)}{r}
	\end{equation}
	where $\alpha$ is a positive constant and $\erf(r)=\frac{2}{\sqrt{\pi}}\int_0^r e^{-u^2}du$ is the error function, so that $U$ is decomposed as
	$U=U_1+U_2$, where
	\begin{equation}
		U_1=\frac{1}{2}\sum_i q_i \left(\sum_j \sum_{\b{n}} q_j \frac{1}{|\b{r}_{ij}+\b{h}\b{n}|}
		\mathrm{erf}\left(\sqrt{\alpha}|\b{r}_{ij}+\b{h}\b{n}|\right)\right)-\sum_i q_i^2 \sqrt{\dfrac{\alpha}{\pi}}
	\end{equation}
	and
	\begin{equation}\label{U2}
		U_2=\frac{1}{2}\sum_{\b{n}}{'}\sum_{ij}  q_i   q_j \frac{1}{|\b{r}_{ij}+\b{h}\b{n}|}
		\mathrm{erfc}\left(\sqrt{\alpha}|\b{r}_{ij}+\b{h}\b{n}|\right).
	\end{equation}
	The sum in $U_2$ now converges absolutely and rapidly, and one can truncate it to simplify the computation. The sum in $U_1$ still converges conditionally in spite of the charge neutrality condition, but since the kernel is smooth, the summation converges rapidly in the Fourier domain. By the Fourier transform, $U_1$ can be rewritten as
	\begin{gather}\label{eq:U1fourier}
		U_1=\frac{2\pi}{V} \sum_{\b{k}\neq \b{0}}\frac{\exp(-|\b{k}|^2/(4\alpha))}{|\b{k}|^2} |\rho(\b{k})|^2-\sum_i q_i^2 \sqrt{\alpha/\pi},
	\end{gather}
	where $\b{k}=2\pi V^{-1/3}\b{m}$ for $\b{m}$ being integer vector in $\mathbb{Z}^3$,
	and
	\begin{equation}
		\rho(\b{k}):=\sum_i q_i e^{i \b{k}\cdot \b{r}_i}.
	\end{equation}

	We need the derivatives of $U$ in two places. One is in the
	equation of $\b{p}_i$, where $\nabla_{\b{r}_i}U$ is needed and the partial derivative is taken by regarding $U$ as a function of $\{\b{r}_i\}$ and $V$.
	The resulted formula is the same as that in \cite{jin2021random}, and given by
	\begin{multline}\label{eq:force}
		\bm{F}_i=-\nabla_{\bm{r}_i}U=-\sum_{\bm{k}\neq \b{0}}\frac{4\pi q_i \bm{k}}{V |\b{k}|^2}
		e^{-|\b{k}|^2/(4\alpha)}\mathrm{Im}(e^{-i\bm{k}\cdot\bm{r}_i}\rho(\bm{k}))\\
		-q_i\sum_{j,\bm{n}}{'} q_j G(|\bm{r}_{ij}+\b{h}\bm{n}|)\frac{\bm{r}_{ij}+\b{h}\bm{n}}{|\bm{r}_{ij}+\b{h}\bm{n}|}=:\bm{F}_{i,1}+\bm{F}_{i,2},
	\end{multline}
	where we recall $\bm{r}_{ij}=\bm{r}_j-\bm{r}_i$, pointing towards particle $j$, and
	\begin{equation}
		G(r):=\frac{\erfc(\sqrt{\alpha}r)}{r^2}+\frac{2\sqrt{\alpha}e^{-\alpha r^2}}{\sqrt{ \pi}r}.
	\end{equation}

	The other place is for the pressure, where we need to compute $-\frac{\partial U}{\partial V}\big{|}_{\b{s}_i}$. For $U_1$ given in Eq.\eqref{eq:U1fourier}, if we regard it as a function of $\{\b{s}_i\}$ and $V$, it is found that $\b{k}\cdot \b{r}_i=2\pi \b{m}\cdot \b{s}_i$ is independent of $V$. Hence, one may compute that
	\begin{equation}
		-\frac{\partial U_1}{\partial V}\bigg|_{\b{s}_i}
		=\frac{2\pi}{V^2}\sum_{\b{k}\neq \b{0}}\frac{|\rho(\b{k})|^2}{|\b{k}|^2}e^{-|\b{k}|^2/(4\alpha)}
		\left(\frac{1}{3}-\frac{|\b{k}|^2}{6\alpha}\right).
	\end{equation}
	For $U_2$, one rewrites $\b{r}_{ij}+\b{h}\b{n}$ as $V^{1/3}(s_i-s_j+\b{n})$. Then, computing the derivative can be performed directly, resulting in
	\begin{equation}
		-\frac{\partial U_2}{\partial V}\bigg|_{\b{s}_i} =\frac{1}{6V}\sum_{\b{n}}{'}\sum_{ij}  F_{ij,\b{n}}\cdot  (\b{r}_{ij}+\b{h}\b{n}),
	\end{equation}
	where
	\begin{equation}
		F_{ij,\bm{n}}= -q_iq_j G(|\bm{r}_{ij}+\b{h}\bm{n}|)\frac{\bm{r}_{ij}+\b{h}\bm{n}}{|\bm{r}_{ij}+\b{h}\bm{n}|}.
	\end{equation}
	
	Hence, the instantaneous pressure is eventually given by
	\begin{multline}\label{eq:pressurecubic}
		P_{\mathrm{ins}}=\frac{1}{3V}\sum_i \frac{|\b{p}_i|^2}{m_i}+\frac{2\pi}{V^2}\sum_{\b{k}\neq \b{0}}\frac{|\rho(\b{k})|^2}{|\b{k}|^2}e^{-|\b{k}|^2/(4\alpha)}
		\left(\frac{1}{3}-\frac{|\b{k}|^2}{6\alpha}\right)\\
		-\frac{1}{6V}\sum_{\b{n}}{'}\sum_{ij}  q_iq_j G(|\bm{r}_{ij}+\b{h}\bm{n}|)|\bm{r}_{ij}+\b{h}\bm{n}| =:P_{1}+P_{2}+P_3.
	\end{multline}
	
	\subsection{The random batch Ewald}
	
	Now, suppose we do the MD simulation using step size $\Delta t$. The Langevin dynamics is discretized using the BAOAB scheme \cite{leimkuhler2013robust}, which has been discussed in Section \ref{discretize}.
	
	We consider the computation of the force and pressure at each time step.  The short range terms $\b{F}_{i,2}$ and $P_{3}$ in Eq.~\eqref{eq:force} and Eq.~\eqref{eq:pressurecubic} respectively can be computed by considering the nearest image only \cite{thompson2009general}, whereas the long range terms $\b{F}_{i,1}$ and $P_{2}$ in Eq.~\eqref{eq:force} and Eq.~\eqref{eq:pressurecubic} respectively are generally the most time-consuming parts in the NPT simulation even when the FFT is employed for acceleration. First note that the factor $e^{-|\b{k}|^2/(4\alpha)}$ can be normalized to a discrete probability distribution \cite{jin2021random}. In fact, denoting the sum of such factors by
	\begin{gather}\label{eq:S}
		S:=\sum_{\bm{k}\neq \b{0}}e^{-|\b{k}|^2/(4\alpha)}=H^3-1,
	\end{gather}
	where
	\begin{gather}
		H:=\sum_{m\in \Z}e^{-\pi^2 m^2/(\alpha L^2)}
		=\sqrt{\dfrac{\alpha L^2}{\pi}}\sum\limits_{m=-\infty}^{\infty}e^{-\alpha m^2L^2}
		\approx\sqrt{\frac{\alpha L^2}{\pi}}(1+2e^{-\alpha L^2}). \label{psf}
	\end{gather}
	Then, one can regard the sum as an expectation over the probability distribution
	\begin{gather}\label{eq:probexpression}
		\mathscr{P}_{\bm{k}}:=S^{-1}e^{-|\b{k}|^2/(4\alpha)},
	\end{gather}
	which, with $\bm{k}\neq \b{0}$, is a discrete Gaussian distribution and may be sampled efficiently. The MD simulations can then be done via the random mini-batch approach with this importance sampling strategy. Specifically, one approximates the force $\bm{F}_{i,1}$ in Eq.\eqref{eq:force} by the following random variable:
	\begin{gather}\label{appforce}
		\bm{F}_{i,1}\approx \bm{F}_{i,1}^*:=-\sum\limits_{\ell=1}^p \dfrac{S}{p}\dfrac{4\pi \bm{k}_\ell q_i}{V k_\ell^2}\mathrm{Im}(e^{-i\bm{k}_\ell\cdot\bm{r}_i}\rho(\bm{k}_\ell)),
	\end{gather}
	and the frequency part of the pressure in Eq.\eqref{eq:pressurecubic}, $P_2$, is analogically approximated as
	\begin{gather}\label{apppress}
		P_2\approx P_2^{*}= \frac{2\pi}{V^2}\sum_{\tilde{\ell}=1}^{\tilde{p}}\frac{S}{\tilde{p}}\frac{|\rho(\b{k}_{\tilde{\ell}})|^2}{|\b{k}_{\tilde{\ell}}|^2}
		\left(\frac{1}{3}-\frac{|\b{k}_{\tilde{\ell}}|^2}{6\alpha}\right)
	\end{gather}
	where $p$ and $\tilde{p}$ are the numbers of $\bm{k}$'s in the two batches (which of course can be the same) for approximating the force and the pressure, respectively. The sampling of $\bm{k}_{\ell}$ is described in the next subsection. We remark that the samples used for the force and the pressure can be different.
	
	\subsection{Sampling the frequencies}\label{sample}
	
	To sample the frequencies where the size of the box is changing, we may use the following strategies:
	\begin{itemize}
		\item (Offline approach) We may sample a series of Gaussian variables $\b{z}_{\ell}$
		offline. Then, for each step, we do $\bar{\b{m}}_{\ell} \leftarrow \b{z}_{\ell}\sqrt{\alpha L^2/(2\pi^2)}$. Then, one may take the integer value of $\bar{\b{m}}_{\ell}$
		to be the $m$ value, and then set $\b{k}_{\ell}=(2\pi/L)\bar{\b{m}}_{\ell}$.
		The offline approach proposed here has systematic errors
		because there is no Metropolis rejection to correct the rounding error.
		It is impossible to sample the discrete $\b{k}_{\ell}$ offline because we do not know $L$ in advance.

		\item (Online approach)  One may also do online approach where the Metropolis rejection step is added to correct the errors. That means in each step when we want to obtain the samples $\b{k}_{\ell}$ or $\b{k}_{\tilde{\ell}}$, we use the $m$ values in the previous step as the starting state to perform the MH algorithm to obtain the $p+p'$ discrete samples $m_j,~j=1,2,3,$ from the discrete distribution $e^{-|m_j|^2\pi^2/(\alpha L^2)}$ for $\vec{m}=(m_1,m_2,m_3)\in \Z^3, \vec{m}\neq \b{0}$.
		Then, set $\b{k}_{\ell}\leftarrow (2\pi/L)\vec{m}$. The proposal in the MH sampling can be chosen to be the continuous Gaussian, because the probability that the continuous Gaussian falls into $I_{m_{j}}:=[m_{j}-1/2, m_{j}+1/2)$ can be computed easily so that the rejection step can be performed efficiently.
		
		\item In practice, we also offer a simple but efficient parallel strategy for sampling. First, assume that $M$ MPI ranks are employed at step 1, and $M$ independent sampling processes are executed in parallel within each rank. Next, at step $j$, $1\leq j\leq M$, the $j$th MPI rank broadcasts the samples to other ranks using block operation. This strategy evaluates and updates the samples every $M$ steps, dramatically reducing the cost of sampling. An additional error will be introduced because all of these $M$ sampling processes employ the information at the initial step, but this error should be small due to the slow variation of the length of the cell. It is validated from numerical experiments that this stategy work well when $M$ is about $10$.
	\end{itemize}
	
	\subsection{Analysis and discussion}
	
	In this subsection, we conduct some analysis and discussion on the algorithm to demonstrate its validity.
	
	Let us first take a look at the complexity of the method.
	Similar to the strategy in the FFT-based method \cite{deserno1998mesh}, we may choose the splitting parameter $\alpha$ such that the computational cost in real space is cheap. The resulting computation cost in the Fourier space is expensive in the usual Ewald sum, but is greatly reduced by the random mini-batch strategy in the RBE method. Referring to \cite{deserno1998mesh}, we make the choice that
	\begin{equation}
		\sqrt{\alpha}\sim\dfrac{N^{1/3}}{L},
	\end{equation}
	which is inverse of the average distance between two numerical particles. The complexities for the real space part and the Fourier part in the usual Ewald sum are $\mathcal{O}(N)$ and $\mathcal{O}(N^2)$, respectively. Thanks to the random batch approximation given in Eq.~\eqref{appforce} and Eq.~\eqref{apppress}, which is some Monte Carlo method for approximating the force and the pressure, the
	number of frequencies considered is then reduced to $\mathcal{O}(p+\tilde{p})$. If we choose the same batch of frequencies for forces acting on different particles  (i.e., using the same $\b{k}_{\ell},~ 1\leq \ell \leq p$ for all $\b{F}_{i,1}^{*}$) in the same time step, the complexity per iteration for the frequency part is reduced
	to $\mathcal{O}((p+\tilde{p})N)$. This implies that the RBE method has linear complexity per time step if one chooses $p = \mathcal{O}(1)$ and $\tilde{p}=\mathcal{O}(1)$.
	
	Define the deviations in the random batch approximation for the Fourier parts of the force and the pressure by
	\begin{equation}\label{eq:forcedevi}
		\b{\chi}_i(\b{r}; V):=\b{F}_{i,1}^{*}(\b{r}; V)-\b{F}_{i,1}(\b{r}; V),
	\end{equation}
	and
	\begin{equation}\label{eq:predevi}
		\tilde{\chi}(\b{r}; V):=P_{2}^{*}(\b{r}; V)-P_{2},
	\end{equation}
	respectively, where $\b{r}=(\b{r}_1, \cdots, \b{r}_N)\in \R^{3N}$. The expectation and variance of the random deviations for fixed configuration $(\b{r}, V)$ can be obtained by direct calculation, which is given by Proposition \ref{expection}.
	
	\begin{proposition}\label{expection}
		Consider some fixed configuration $(\b{r}, V)$.	The deviations in force $\b{\chi}_i$ and pressure $\tilde{\chi}$ have zero expectation
		\begin{equation}\label{EE}
			\mathbb{E}\b{\chi}_i=0,~~~\mathbb{E}\tilde{\chi}=0,
		\end{equation}
		and that the variances are
		\begin{equation}\label{ForceExp}
			\mathbb{E}\left|\boldsymbol{\chi}_{i}\right|^{2}=\frac{1}{p}\left(\sum_{\boldsymbol{k} \neq \b{0}} \frac{\left(4 \pi q_{i}\right)^{2} S}{V^{2} |\b{k}|^{2}} e^{-\frac{|\b{k}|^{2}}{4 \alpha}}\left|\operatorname{Im}\left(e^{-i \boldsymbol{k} \cdot \boldsymbol{r}_{i}} \rho(\boldsymbol{k})\right)\right|^{2}-\left|\boldsymbol{F}_{i, 1}\right|^{2}\right),
		\end{equation}
		and
		\begin{equation}\label{PressExp}
			\mathbb{E}|\tilde{\chi}|^2=\frac{1}{\tilde{p}}\left(\sum_{\b{k} \neq \b{0}} \frac{4\pi^{2}|\rho(\boldsymbol{k})|^{4}S}{V^{4} k^{4}} e^{-\frac{|\b{k}|^{2}}{4 \alpha}}\left(\dfrac{1}{3}-\dfrac{|\b{k}|^2}{6\alpha}\right)^{2}-\left|P_2\right|^2\right).
		\end{equation}
	\end{proposition}
	
	The claims, Eq.~\eqref{EE}, in Proposition \ref{expection} imply that the random approximation is consistent or unbiased,
	\begin{equation}
		\mathbb{E}\b{F}_{i,1}^{*}=\b{F}_{i,1},~~~\mathbb{E}P_{2}^{*}=P_{2},
	\end{equation}
	where $\mathbb{E}$ is the expectation over the random batches. The other two claims, Eq.\eqref{ForceExp} and Eq.\eqref{PressExp}, say that the variances of the approximate force $\b{F}_{i,1}^{*}$ and the approximate pressure $P_{2}^{*}$ are like $\cO(p^{-1})$ and $\cO(\tilde{p}^{-1})$, respectively.
	
	By the expressions \eqref{eq:forcedevi} and \eqref{eq:predevi}, the deviations in the random approximation are of $\cO(1)$ for each step. At the first glance, this seems unacceptable as this error is big. The rationalization of the method is that as the dynamics goes on, the random approximations accumulate in time and \eqref{EE} tells us that the averaged effect is correct. Hence the random batch type methods work due to this time averaging effect which can be regarded as the law of large number in time. This type of Monte Carlo method can have the
	strong error like $\sqrt{\mathrm{variance}*\Delta t}$.
	In earlier works \cite{jin2020random, jin2021convergence}, the strong error estimates for random batch type methods have been established rigorously for some regular potentials.
	The error estimates are expressed as,
	\begin{gather}\label{eq:errorbound}
		\left(\E\frac{1}{N}\sum_{i=1}^N|X_i-\tilde{X}_i|^2\right)^{1/2}\lesssim \sqrt{\Lambda(N) \Delta t},
	\end{gather}
	where $X_i$ represents the locations and quantities for the particles to be computed (corresponding to $\b{r}_i, \b{p}_i$, $V$ and $p^V$ in our case), and $\Lambda(N)$ is the upper bound for the variance of the random approximation. In the mean field regime, $\Lambda(N)$ is independent of $N$. In our case, the pressure and the forces are singular at $\b{r}_i=\b{r}_j$ for some $i\neq j$. The rigorous justification is challenging.
	Neverthless, we expect that the error bound \eqref{eq:errorbound} still holds. Recalling \eqref{eq:probexpression}, the Fourier modes with low frequency are more likely to be chosen in the random mini-batch approach.  Since the long wave modes are more
	important for the periodic effects, this importance sampling strategy could be more
	effective compared with the uniform sampling across the modes considered.
	With this importance sampling strategy, the variance can be reduced so that the random method is more accurate.
	In fact, as we can see  in \eqref{ForceExp} and \eqref{PressExp}, the variances are insensitive to the particle number $N$ and thus the random batch Ewald method is expected to be effective. This will be justified in the numerical tests in section \ref{sec:numerics}.

\section{Simulation results}\label{sec:numerics}

In this section, we perform several all-atom numerical results with the RBE-based MD under the NPT ensemble for the cubic cell (SPC/E bulk water systems) and the rectangular cell with semi-isotropic coupling along the $xy$ plane (DPPC membrane systems) to validate the performance of the method. Without loss of generality, $p$ and $\tilde{p}$, the numbers of $\bm{k}$'s in a batch for approximating the force and the pressure, are chosen as the same in all tests for convenience.

\subsection{Accuracy benchmark on bulk water}
We first test the Langevin equations combined with the RBE for NPT on the simulation of bulk water. The system includes $17789$ SPC/E water molecules confined in a cubic box of initial side length $8.16$ nm. For each case, a short equilibration run of $200$ ps is first conducted in the NVT ensemble, at reference pressure of $10^{-3}$ Kbar and reference temperature $298$ K,  with the integration step size $\Delta t=1$ fs. The relaxation times are set to $\gamma_i$ = $0.1$ ps for each particle $i$ and $\gamma$ = $0.5$ ps for the cell. The production phase lasts $2$ ns, and the configurations are saved every $200$ steps ($0.2$ ps) for statistics. The initial guess of the compressibility takes $4.5 \times 10^{-2}$ Kbar$^{-1}$ (which is actually a value taken from the liquid water under ambient condition). These simulations were conducted by our method implemented in the LAMMPS \cite{plimpton1995fast,thompson2021lammps} (version 29Oct2020). LAMMPS also includes a stable implementation of Hoover type equations of motion (assembled in the ``fix npt'' command), comprising both a modified version \cite{shinoda2004rapid} of the Martyna-Tobias-Klein \cite{martyna1994constant} (MTK) algorithm combined the hydrostatic equations with the strain energy \cite{parrinello1981polymorphic} and a measure-preserving time integrator \cite{tuckerman2006liouville}, so that we can compare it to our method. In some figures and their captions in this paper, this ``fix npt'' equations of motion is marked as ``MMTK''.

\begin{figure}[htbp]
	\centering
	\includegraphics[width=1.0\linewidth]{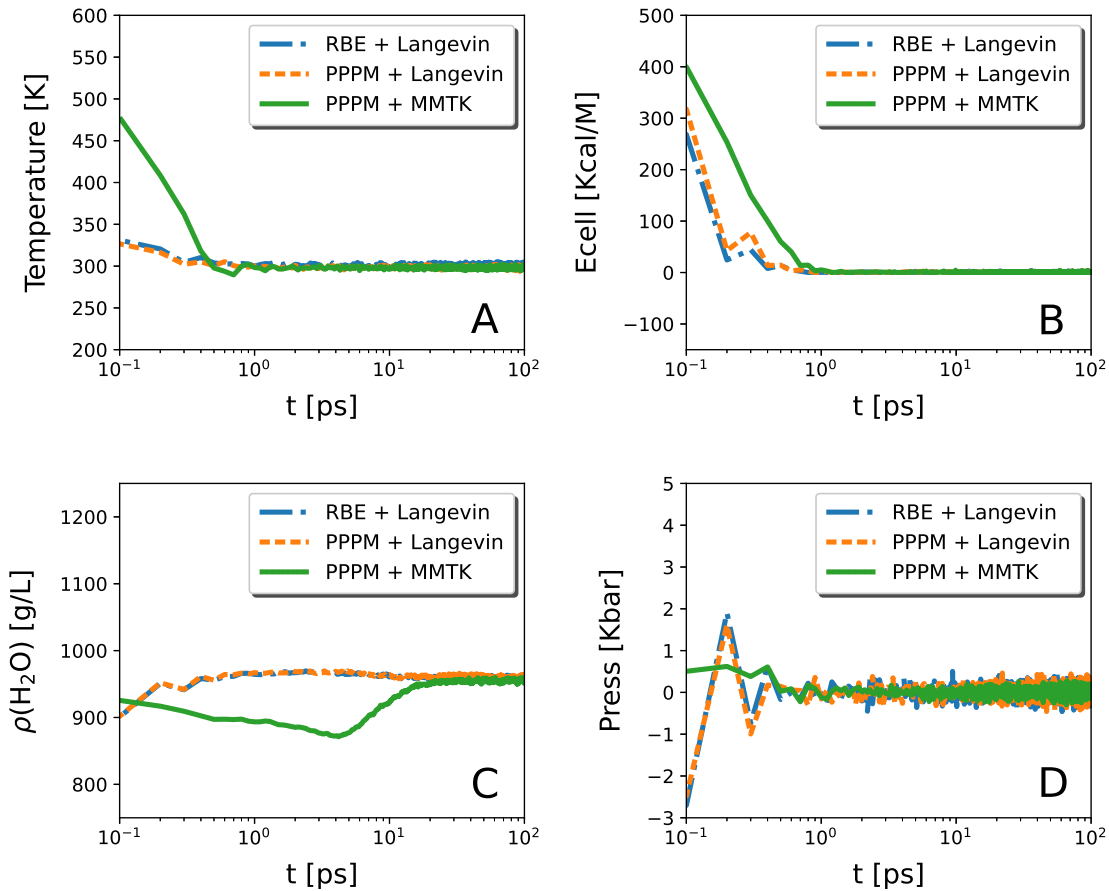}
	\caption{The equilibration of the bulk water system toward 298 K and $10^{-3}$ Kbar. The simulation starts from a nonequilibrium configuration at $0$ K. The first $100$ ps is captured for statistics. The plots present the equilibration of (A) temperature, (B) $E_{\text{cell}}$, (C) $\rho$($\text{H}_2$O), and (D) presssure combined with different electrostatic solvers (blue dash-dotted line for the RBE and orange dashed line for the PPPM). The green solid lines show results of the reference simulations produced via the MMTK equations of motion provided in LAMMPS combined with the PPPM.}
	\label{fig:1_1}
\end{figure}

To assess the performance of our method, we first study the equilibrium time of four physical quantities: temperature, pressure, kinetic energy of the simulation box $E_{\text{cell}}$, and density of water molecules $\rho(\text{H}_{2}\text{O})$. Fig.~\ref{fig:1_1} (A-D) shows that the system is successfully equilibrated to the desired thermodynamic state within only $0.6$ ps (observably faster than the reference), for both the RBE (with $p=200$) and the PPPM conducted with the Langevin equations of motion. The fluctuations of all these four quantities are relatively small even for the RBE with small batches.

\begin{figure}[htbp]
	\centering
	\includegraphics[width=1.0\linewidth]{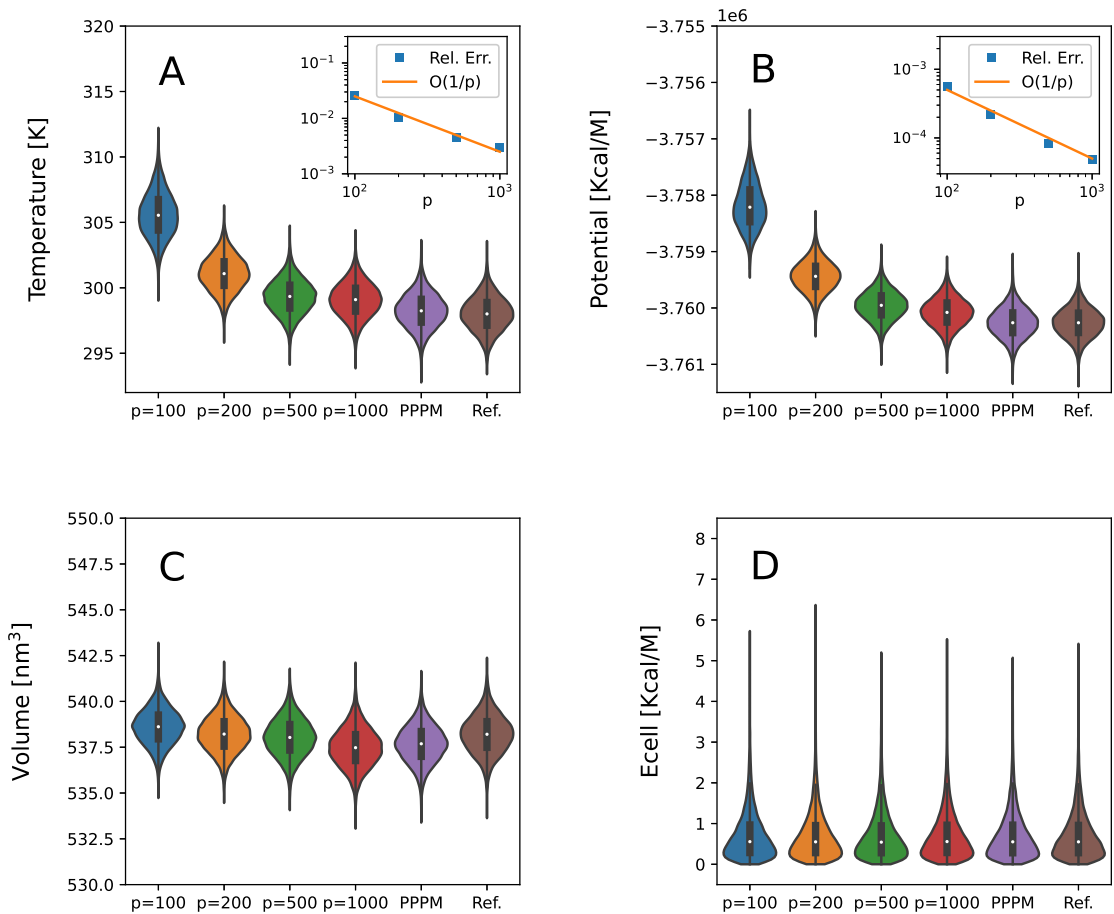}
	\caption{The violin plots of temperature (A), total potential energy (B), volume (C), and $E_{\text{cell}}$ (D). Data are shown for the RBE with different choices of $p$, the numbers of $\bm{k}$'s in a batch (blue for $p=100$, yellow for $p=200$, green for $p=500$, and red for $p=1000$), and the PPPM (violet) combined with the Langevin integrator. The white point and the two endpoints of black bar within each violin indicate the mean value and two quartiles, respectively. The subfigures in A and B show the convergences on the mean values of both temperature and total potential energy with $O(p^{-1})$ rate. The statistical distributions of these four quantities are also observed to be consistent with the ``Ref.'' results, which can be seen as a reference produced by the PPPM combined with the MMTK integrator in LAMMPS. }
	\label{fig:1_2}
\end{figure}

We then evaluate the statistical robustness of the equations by measuring averages and fluctuations of temperature, total potential energy, volume, and $E_{\text{cell}}$. The time-average distributions of these four quantities are plotted in Fig.~\ref{fig:1_2}. In the figures, the results of both the distributions and mean values of the NPT Langevin dynamics are given, illustrating the consistency with the reference simulations. The distributions of Fig.~\ref{fig:1_2} (A-C) present the desired Boltzmann distribution, further confirming the correctness of both the equations and our numerical approximations. In the subplot of Fig.~\ref{fig:1_2} (A-B), the convergence of both temperature and total potential energy shows the $O(p^{-1})$ rate in consistent with our priori error estimate.

\begin{figure}[!htbp]
	\centering
	\includegraphics[width=1.0\linewidth]{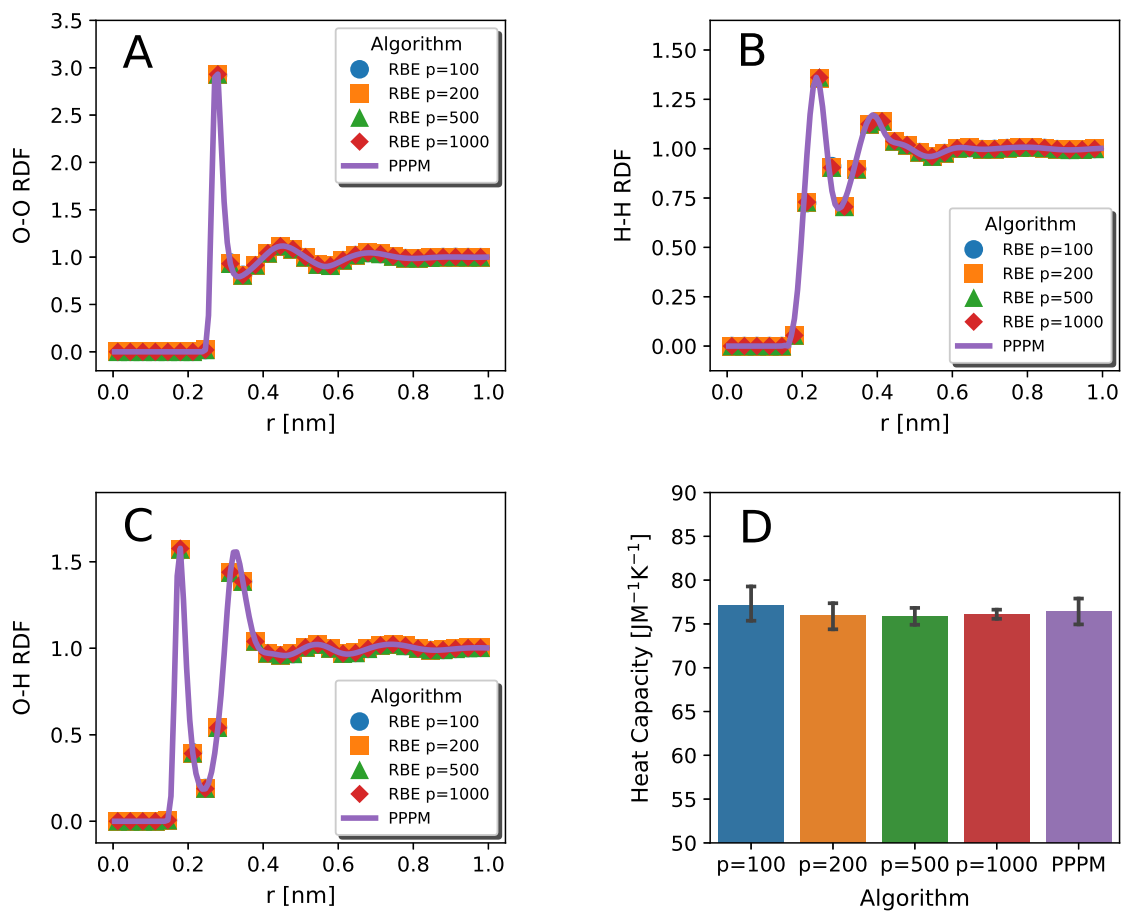}
	\caption{The RDFs of O-O (A), H-H (B), and O-H (C) and heat capacity (D) of the bulk water system. The simulation results are plotted for the PPPM (violet in (A-D)) and the RBE with different numbers of $\bm{k}$'s in a batch, $p$, including $p=100$ (blue circle), $p=200$ (orange square), $p=500$ (green triangle), and $p=1000$ (red rhombus) in(A-C) (and distinguished by corresponding colors in (D)). Datas produced by the RBE are also combined with the Langevin equations of motion developed in this paper. Datas produced by the PPPM are combined with the MMTK equations of motion. All the results show great agreements with the reference, demonstrated the ability of the RBE on producing accurate structural properties.}
	\label{fig:1_3}
\end{figure}

\begin{figure}[htbp]
	\centering
	\includegraphics[width=1.0\linewidth]{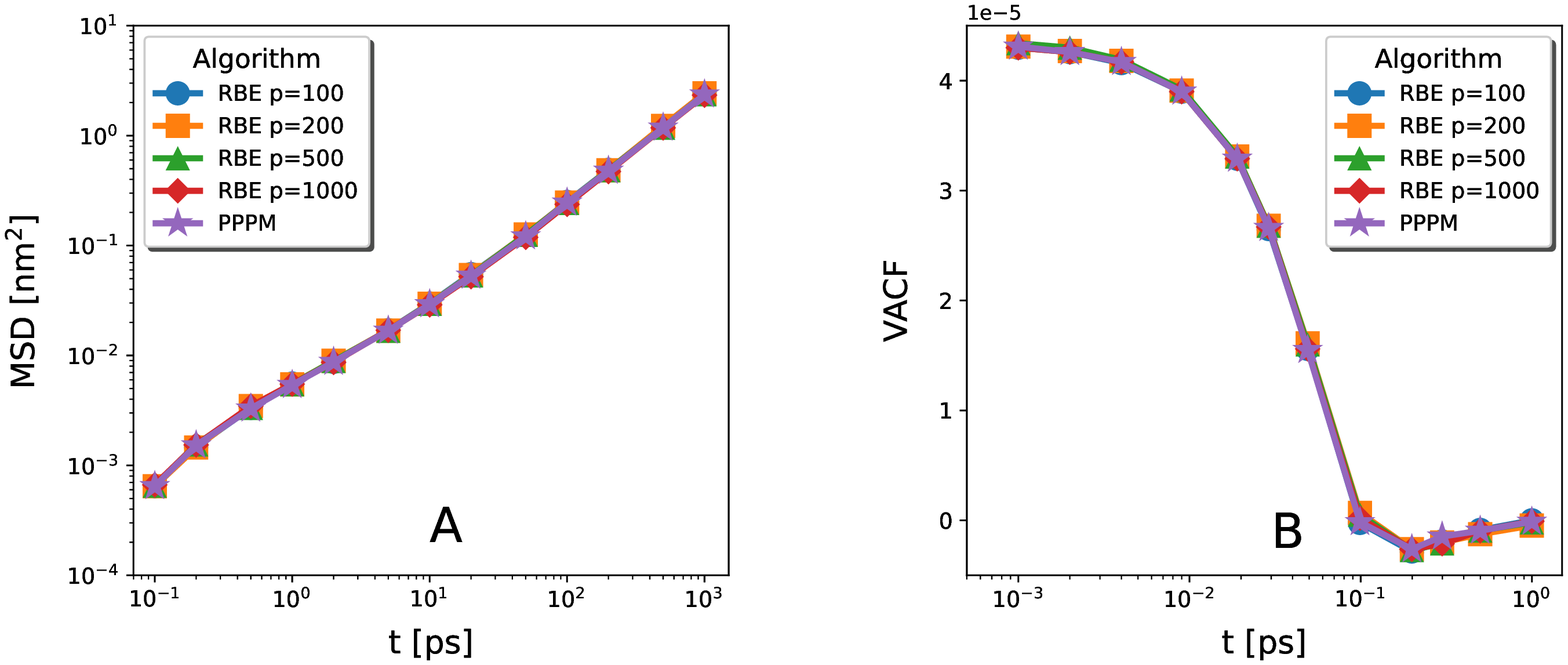}
	\caption{The MSD (A) and the VACF (B) of the bulk water system. The simulation results are plotted for the PPPM (violet five-pointed star) and the RBE with different $p$, the numbers of $\bm{k}$'s in a batch, including $p=100$ (blue circle), $p=200$ (orange square), $p=500$ (green triangle), and $p=1000$ (red rhombus). The accuracy of the RBE on producing accurate dynamic information is verified.}
	\label{fig:1_4}
\end{figure}

Next, we examine the accuracy of the RBE-based NPT simulation for the water system using the Langevin equations of motion, compared to the PPPM. We calculate the radial distribution (RDF), the isobaric heat capacity, the mean-squared displacement (MSD), and the velocity auto correlation function (VACF) of water molecules. The RDF of oxygen-oxygen (O-O), hydrogen-hydrogen (H-H), and oxygen-hydrogen (O-H) atom pairs furnish the spatial arrangement of water molecules. The isobaric heat capacity describes the heat exchanged between the system and the environment under NPT ensemble. The MSD of oxygen describes the translation motion on wide range of time scales. The VACF characterizes the short-time dynamics of water. The comparisons on the structure and the thermodynamic properties, including the RDFs of O-O, H-H, and O-H and the isobaric heat capacity, are displayed in Fig.~\ref{fig:1_3} (A-D), respectively. The RBE and the PPPM integrated with the Langevin equations produce statistically identical results on all of these three RDFs, which are in agreement with the reference solution and the literature result \cite{mark2001structure}. The convergence on the isobaric heat capacity with the increase of $p$ is also observed, which demonstrates that the RBE with $p=200$ produces approximately identical results as the PPPM. Fig.~\ref{fig:1_4} shows the comparisons on the dynamical properties. The perfect agreement between the RBE and the PPPM confirms that dynamical properties are properly reproduced, indicating the correct integration of the RBE with the Langevin equations of motion. Note that the reference ``exact" solution employing existing method is not provided in Fig.~\ref{fig:1_4}, because different integral methods will normally generate different dynamics but produce the same structural properties.

\begin{figure}[htbp]
	\centering
	\includegraphics[width=1.0\linewidth]{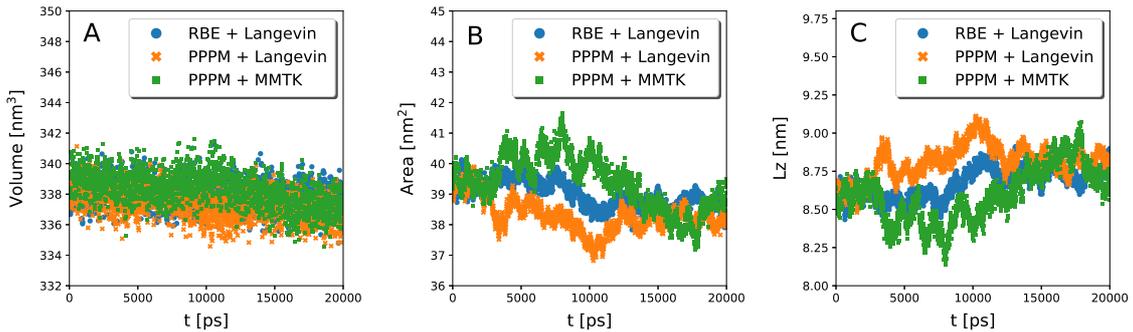}
	\caption{Time series from the membrane system (DPPC) simulations. (A) Volume, (B) area of the simulation box in the xy plane, and (C) height of the simulation box along the z axis. Data are shown for the reference simulation (using the MMTK equations of motion in LAMMPS combined with the PPPM, in green square) and the Langevin equations of motion developed in this paper combined with the RBE (using $p=500$, in blue circle) and the PPPM (setting the estimated relative error $\Delta=10^{-4}$, in orange times sign). The real-space cutoff is set to $1.0$ nm for both the RBE and the PPPM.  }
	\label{fig:1_6}
\end{figure}

\subsection{Accuracy benchmark on membrane simulation}
In the calculations, the membrane system is built by the CHARMM-GUI \cite{jo2008charmm}, with force field CHARMM36m \cite{huang2017charmm36m}. After solvation, a system composed of $140$ dipalmitoylphosphatidylcholine (DPPC) lipid and $5801$ SPC water molecules \cite{mark2001structure} is obtained.

For this example, we apply the semi-isotropic fluctuation equations given in Appendix \ref{extensions}. The random batch approximations of the forces and pressure can be computed similarly as the cubic boxes except that the pressures along $x$ and $y$ directions are coupled for this membrane simulation.
Similar to the bulk water system, the production phase consists of two sets of NPT simulations using the ``fix npt'' command in LAMMPS, which acts as the reference simulation. And we simulate the Langevin barostats developed in this paper by combining it with the RBE and the PPPM, respectively.
A reference temperature of $303.15$ K, a reference pressure
of $10^{-3}$ Kbar, and a surface tension of $0$ dyn cm$^{-1}$ are employed for semi-isotropic pressure coupling (i.e., $x$ and $y$ axes are coupled in computing the virial), along with thermostat relaxation time $\gamma_i=0.1$ ps, barostat relaxation time $\gamma$ = $0.5$ ps and a compressibility of $4.5\times10^{-2}$ Kbar$^{-1}$. All simulations run for $40$ ns and configurations for every 5000 steps ($5$ ps) are saved for statistics.
All shown results were obtained by analyzing the second half of the trajectory. In this example, the temperature coupling is performed on two groups of atoms separately, where the first group comprises the protein and DPPC lipids and the second one includes the solvent and ion molecules. This is because energy exchange between different components is not perfect and should be investigated carefully \cite{eastwood2010equipartition}, due to different effects including cut-offs etc.

We test the effect of the electrostatic algorithm on the equilibration of a model membrane. Time series for the simulated trajectories are reported in Fig.~\ref{fig:1_6}. It is clearly shown that the Langevin barostat combined with the RBE or the PPPM provides consistent results with the reference solution (produced by the MMTK combined with the PPPM). Note that the DPPC simulations exhibit natural fluctuations in the surface area with a standard deviation of about $3\%$, and all of our simulations yield fluctuations less than this value. The mean value and the standard deviation are given in Tab.~\ref{Table2}, where we can see that the fluctuations of the RBE on standard deviation are slightly smaller than the PPPM. This result indicates that the RBE with $p=500$ has comparable accuracy compared with the PPPM with $\Delta=10^{-4}$ in this DPPC system.

\renewcommand\arraystretch{1.6}
\begin{table*}[!htbp]
	\centering
	\setlength{\tabcolsep}{2.8mm}{
		\begin{tabular}{c|c|c|c}	
			& RBE+Langevin & PPPM+Langevin & PPPM+MMTK
			\\\hline
			$\langle$Volume$\rangle$ $[nm^3]$ & 338.002 & 337.346 & 338.305  \\\hline
			Std. Volume $[nm^6]$ & 0.853 & 0.950 & 1.091  \\ \hline
			$\langle$Area$\rangle$ $[nm^2]$ & 38.972 & 38.307 & 39.194  \\ \hline
			Std. Area $[nm^4]$& 0.455 & 0.534 & 0.865  \\ \hline
			$\langle$Lz$\rangle$ $[nm]$ & 8.674 & 8.808 & 8.636  \\ \hline
			Std. Lz $[nm^2]$& 0.097 & 0.112 & 0.177  \\ \hline
	\end{tabular}}
	\caption{Average and fluctuations of the volume, area of the simulation box along the xy plane (Area) and length of the box along the $z$ axis ($L_z$) for the DPPC membrane system. Data are produced by the same simulations in Fig.~\ref{fig:1_6}.}
	\label{Table2}
\end{table*}

\subsection{Time performance of the RBE-based NPT}
The performance comparison between the RBE-based and the PPPM-based NPT simulation is carried out by using the MD engine of LAMMPS on the same SPC/E bulk water system comprising $17789$ water molecules. The estimated relative force error $\Delta$ is chosen as $10^{-4}$. The parameters of the PPPM are chosen automatically in LAMMPS based on the error estimates \cite{deserno1998mesh}. The real space cutoff $r_c$ is set to be $10$ $\mathring{A}$ for both the PPPM and the RBE. The simulations of the system are conducted for 20000 steps to estimate the average CPU time per step. The simulations are performed on the $\pi$ 2.0 cluster at the Center for High Performance Computing of Shanghai Jiao Tong University, for which each CPU node contains two Intel Xeon Scalable Cascade Lake 6248 (2.5GHz, 20 cores) and 12 $\times$ Samsung 16GB DDR4 ECC REG 2666 memory.

\begin{figure}[htbp]
	\centering
	\includegraphics[width=1.0\linewidth]{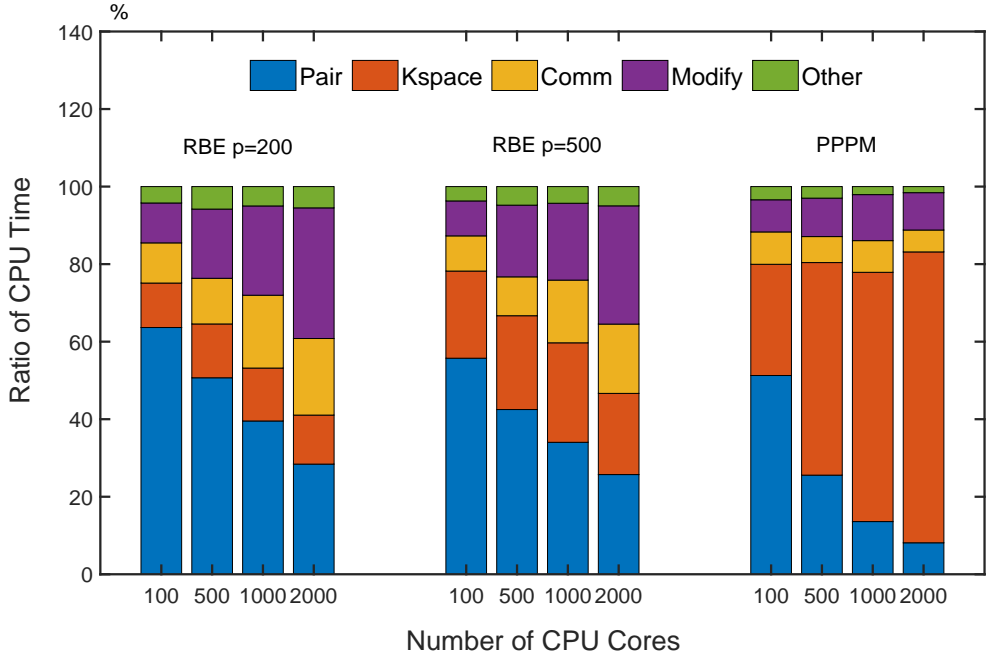}
	\caption{The percentages of the total CPU time per step that are spent on each part, including \emph{Pair} (blue), \emph{Kspace} (red), \emph{Comm} (yellow), \emph{Modify} (violet), and \emph{Other} (green). The specific explanations of these parts are described in the main paper. On the whole, the time-consuming of the \emph{Pair} part with respect to a fixed number of cores is roughly identical for both the RBE and the PPPM. It is observed that the RBE significantly reduces the cost on the Fourier space compared with simulation produced the PPPM, as well as the CPU time for the whole NPT simulation (listed in Tab. \ref{Table1}).}
	\label{fig:1_5}
\end{figure}

Fig.~\ref{fig:1_5} shows the percentages of the total CPU time per step that are spent on four components including \emph{Pair}, \emph{Kspace}, \emph{Comm}, \emph{Modify}, and \emph{Other}, with varying CPU cores. The quantity \emph{Pair} is the sum of CPU time of all non-bonded force computation, including the Lennard-Jones(LJ) force and the near-field part of the Coulomb force. \emph{Kspace} is an important quantity, containing CPU time of the long-range part of the Coulomb interactions. \emph{Comm} records the inter-processor communication cost of atoms and their properties, e.g., the inverse communication when one employs Newton's third law of motion \cite{plimpton1995fast}. \emph{Modify} counts the cost of ``fixes" and ``computes" operations in LAMMPS, including the time integration, the thermostat, the barostat, and the SHAKE algorithm \cite{krautler2001fast} employed for the constraints of all chemical bonds. Cost for other remaining part, including the bonded interactions, the neighbor list constructions, {\it et al.}, are summarized in the \emph{Other} quantity for convenience (note that in the classical LAMMPS, these parts can be counted separately).

\renewcommand\arraystretch{1.6}
\begin{table*}[!htbp]
	\centering
	\setlength{\tabcolsep}{1.9mm}{
		\begin{tabular}{|c|cc|cc|cc|cc|}	
			\hline \multirow{3}{*}{} &
			\multicolumn{8}{c|}{Number of CPU Cores}
			\\\cline{2-9}
			& \multicolumn{2}{c|}{100} & \multicolumn{2}{c|}{500} & \multicolumn{2}{c|}{1000} & \multicolumn{2}{c|}{2000}
			\\\cline{2-9}
			&Kspace &Total&Kspace &Total&Kspace &Total&Kspace &Total
			\\\hline RBE P=200 &2.20e-4  & 1.92e-3 &8.27e-5  &5.96e-4
			&5.87e-5 & 4.30e-4&4.76e-5 & 3.77e-4\\\hline
			RBE P=500 &4.92e-4  &2.19e-3  &1.75e-4  &7.24e-4  &1.20e-4 &4.99e-4 & 8.71e-5 & 4.16e-4\\\hline
			PPPM &6.83e-4 & 2.38e-3 &5.83e-4 & 1.16e-3 &7.82e-4 & 1.22e-3 &9.88e-4 & 1.32e-3 \\ \hline
	\end{tabular}}
	\caption{Average total CPU time per step (units: seconds) as a function of different number of CPU cores produced by the RBE and the PPPM for the simulations in Fig.~\ref{fig:1_5}.}
	\label{Table1}
\end{table*}

Note that the real CPU costs of \emph{Pair}, \emph{Comm}, \emph{Modify}, and \emph{Other} with respect to a fixed number of core are roughly identical for both the RBE (with $p=200$ or $p=500$) and the PPPM, because they are not affected by the long-range interaction. It is obvious that the percentage of CPU cost occupied by the PPPM significantly increases along with the number of CPU cores with respect to the RBE, as shown in Fig.~\ref{fig:1_5}. This is largely because the required intensive communications of the PPPM, whereas only one global operation per step is needed for the RBE. When $2000$ CPU cores is employed, the CPU time of \emph{Kspace} of the RBE-based simulation has more than one magnitude faster than the PPPM-based. This is important because \emph{Kspace} is generally the bottleneck in large scale MD simulation on massive supercomputing.

In other words, for the total CPU cost of NPT ensemble, the performance advantages of the RBE-based simulation has some reductions compared with the advantages on \emph{Kspace} part, as the results shown in Fig.~\ref{fig:1_5} and Tab.~\ref{Table1}. This phenomenon appears when the CPU cost is dominated by other imperceptible operations (\emph{Comm}, \emph{Modify}, and \emph{Other} in Fig.~\ref{fig:1_5}), including thermostat, bond angle, construction of neighbor list, data statistics, timekeeping and diagnostic routine, which have different requirements in various systems. Generally, costs of these parts are not obvious but will have considerable impact on top of the acceleration of long-range interactions. We look forward to optimize these components in mainstream packages which may become bottleneck in the future.

\section{Conclusions} \label{conclusion}
We have developed a new RBE algorithm that is accurate and efficient for the MD simulaions under NPT ensemble by integrating the global stochastic underdamped Langevin dynamics, derived using a suitable Lagrangian for the NPT ensemble. The method builds the random mini-batch strategy into the Fourier space in the Ewald summation for the pressure and forces so that it takes $\mathcal{O}(N)$ per time step.
The simulations on bulk water and membrane systems show that the RBE algorithm can quantitatively reproduce the spatiotemporal information and thermodynamic quantities, and shows  attractive performance regarding the efficiency and scalability on massive supercomputer cluster. 
These indicate that the RBE algorithms for NPT ensemble can be useful, reliable, and cost-effective for large-scale MD simulations on modern computer architecture.

\section*{Acknowledgements}
 All the authors are supported by the Shanghai Science and Technology Commission (grants 20JC1414100 and 21JC1403700). J. L., and Z. X. are supported by the NSFC (grant 12071288). P. T. and L. H. are supported by NSFC (grants 11974239 and 31630002), the Innovation Program of Shanghai Municipal Education Commission, and the Shanghai Jiao Tong University Multidisciplinary Research Fund of Medicine and Engineering (YG 2016QN13). S. J. is supported by NSFC 12031013. L. L. is supported by NSFC 11901389 and NSFC 12031013. The authors acknowledge the Center for High Performance Computing at Shanghai Jiao Tong University for the computing resources.
	
\appendix

\section{Extensions to general cell fluctuations under Langevin dynamics}\label{extensions}

In this section, we consider general cases, including fully flexible cell and rectangular cell with anisotropic fluctuation. The Langevin equations for these extended cases can be derived similarly. In fact, the Langevin equations for the fully flexible cells have been investigated already in \cite{feller1995constant,quigley2004langevin,gao2016sampling}, and our derivation above for the isotropic case and the semi-isotropic case are motivated by these works.

The reduced variables $\b{s}_i$ satisfies
\begin{equation}
	\b{r}_i=\b{h}\b{s}_i=\sum_{j=1}^3 s_i^{(j)}\b{h}_{j},
\end{equation}
similar derivation can be performed for the rectangular cell with anisotropic fluctuation and the fully flexible cells.

\subsection*{Equations of motion for semi-isotropic case}

Recall the distribution function of constant surface-tension ensemble $NP\gamma_0T$ given in Eq.\eqref{semiisotropic}
\begin{equation}
	g(\b{r}, \b{p}, A, L)\propto \exp(-\beta(K+U+PAL-\gamma_0A)),
\end{equation}
where $K,U$ are the kinetic and potential energies for the particles, $A$ is the area of the simulation box in the $xy$ plane, $L$ is its height, and $\gamma_0$ is the surface tension multiplied by the number of surfaces. The partition function has simple expression
\begin{equation}
	\Delta=\int_{[0,\infty)\times[0,\infty)}dAdL\exp(-\beta PAL)Q(A,L),
\end{equation}
where
\begin{equation}
	Q(A,L)=\int_{D(A,L)}\exp(-\beta (K+U-\gamma_0 A))d\b{r}_id\b{p}_i
\end{equation}
is the partition function for the canonical ensemble with the region fixed to be $D(A,L)$.

Without loss of generality, we assume that the initial simulation cell has square surface in the $xy$ plane. Consider Lagrangian of the following form
\begin{equation}
	\mathscr{L}(\b{s}_i, \dot{\b{s}}_i, A, \dot{A}, L, \dot{L})
	=\sum_{i=1}^N \frac{1}{2}m_i |\b{h}\dot{\b{s}}_i|^2
	+\frac{1}{2}M_A\dot{A}^2+\frac{1}{2}M_L\dot{L}^2 -\left(U(\{\b{h} \b{s}_i\}; A;L) + PAL-\gamma_0 A\right).
\end{equation}
where $\b{h}=\text{diag}(\sqrt{A},\sqrt{A},L)$, so that one obtains the Hamiltonian
\begin{equation}
	\mathscr{H}(\b{s}_i, \b{p}^s_i, A, p^A, L,p^L)
	=\sum_{i=1}^N \frac{|\b{h}^{-1}\b{p}_i^s|^2}{2m_i}
	+\frac{(p^A)^2}{2M_A}+\frac{(p^L)^2}{2M_L}+U(\{\b{h}\b{s}_i\}; A;L) + PAL-\gamma_0A.
\end{equation}

The instantaneous pressure components can be computed similarly by
\begin{gather}
	\begin{split}
		& P_{\mathrm{ins},A}=-\frac{1}{L}\frac{\partial E}{\partial A}\Big|_{s_i,L}
		=\dfrac{P_{\text{ins},xx}+P_{\text{ins},yy}}{2}+\frac{\gamma_0}{L}, \\
		& P_{\text{ins},L}=-\dfrac{1}{A}\dfrac{\partial E}{\partial L}\Big|_{s_i, L}=P_{\text{ins},zz},
	\end{split}
\end{gather}
where
\[
\dfrac{P_{\text{ins},xx}+P_{\text{ins},yy}}{2}=-\dfrac{1}{L}\left(\dfrac{\partial K}{\partial A}+\dfrac{\partial U}{\partial A}\right),~~~P_{\text{ins},zz}=-\dfrac{1}{A}\left(\dfrac{\partial K}{\partial L}+\dfrac{\partial U}{\partial L}\right),
\]
are the components of the interal pressure due to the particles. These relationships  have also been mentioned in the Supplementary Material of \cite{bernetti2020pressure}, whereas we note that there are two symbol typos in Eq.~(S19a) and Eq.~(S19b) of \cite{bernetti2020pressure}.

The general Langevin equations Eq.\eqref{eq:Langevingeneral} in this case can then be written as
\begin{gather}\label{99999}
	\begin{split}
		& \dot{\b{r}}_i=\frac{\b{p}_i}{m_i}+\text{diag}\left(\frac{\dot{A}}{2A},\frac{\dot{A}}{2A},\frac{\dot{L}}{L}\right) \b{r}_i,\\
		&\dot{\b{p}}_{i}= -\nabla_{\b{r}_i}U-\text{diag}\left(\frac{\dot{A}}{2A},\frac{\dot{A}}{2A},\frac{\dot{L}}{L}\right)\b{p}_i
		-\gamma_i \b{p}_i+\sqrt{2k_{\text{B}}Tm_i\gamma_i} \dot{\b{W}}_i,\\
		& \dot{A}=\frac{p^A}{M_A},\\
		&\dot{p}^A=L(P_{\text{ins},A}-P)-\gamma(A)p^A +\sqrt{2k_BT\gamma(A) M_A} \dot{W}_{A},\\
		& \dot{L}=\frac{p^L}{M_L},\\
		&\dot{p}^L=A\left(P_{\text{ins},zz}-P\right)-\gamma(L)p^L +\sqrt{2k_BT\gamma(L) M_L} \dot{W}_{L}.
	\end{split}
\end{gather}

Same as what we discussed in the isotropic case, the differential equation for the constant surface-tension ensemble of the cell rescaling (Eq.~(9a) and Eq.~(9b) in Ref.~\cite{bernetti2020pressure}) is an asymptotic case of the Langevin equations Eq.~\eqref{99999}. Following the
Smoluchowski-Kramers approximation given in \cite[Theorem 1]{hottovy2015smoluchowski},  we consider the equations for $(A,p^{A})$ as the example. Fixing $\tilde{\gamma}(A):=\gamma(A)M_A$ as $M_A\rightarrow 0$, one then obtains the well-known over-damped Langevin dynamics
\begin{equation}
	\dot{A}=\dfrac{L}{\tilde{\gamma}(A)}\left(\dfrac{P_{\text{ins},xx}+P_{\text{ins},yy}}{2}-P+\frac{\gamma_0}{L}\right)-\dfrac{k_{\text{B}}T}{\tilde{\gamma}^2(A)}\dfrac{d \tilde{\gamma}(A)}{d A}+\sqrt{\dfrac{2k_{\text{B}}T}{\tilde{\gamma}(A)}}\dot{W}_{A}.
\end{equation}
If we take $\varepsilon_A=\log(A)$ and $\tilde{\gamma}(A)=\dfrac{3\tau_P L}{2\beta_TA}$, where $\beta_{T}$ is an estimate of the isothermal compressibility of the system and $\tau_{P}$ is a characteristic time associated to the barostat, one obtains
\begin{equation}\label{11a}
	\dot{\varepsilon}_A=\dfrac{2\beta_{T}}{3\tau_{P}}\left(\dfrac{P_{\text{ins},xx}+P_{\text{ins},yy}}{2}-P+\frac{\gamma_0}{L}\right)+\sqrt{\dfrac{4k_{\text{B}}T\beta_{T}}{3AL\tau_{P}}}\dot{W}_{A}.
\end{equation}

The over-damped Langevin dynamics of the conjugate variables $(L,p^{L})$ can be similarly derived as
\begin{equation}\label{11b}
	\dot{\varepsilon}_L=\dfrac{\beta_{T}}{3\tau_{P}}\left(P_{\text{ins},zz}-P\right)+\sqrt{\dfrac{2k_{\text{B}}T\beta_{T}}{3AL\tau_{P}}}\dot{W}_{L},
\end{equation}
where again $\varepsilon_L=\log(L)$.  Eq.~\eqref{11a} and Eq.~\eqref{11b} are exactly the Eq.~(9a) and the Eq.~(9b) in Ref.~\cite{bernetti2020pressure}, respectively.

\subsubsection*{Rectangular cell with anisotropic cell fluctuation}

The Hamiltonian for the rectangular cell with anisotropic cell fluctuation described by the three side lengths $\ell_{\alpha}$ is given by
\begin{gather}
	\mathscr{H}(\b{s}_i, \b{p}^s_i, \ell_{j}, p^{\ell}_{j})
	=\sum_{i=1}^N \frac{|\b{h}^{-1}\b{p}^s_i|^2}{2m_i}+\sum_{j=1}^3\frac{(p^{\ell}_{j})^2}{2M_{j}}
	+\Big(U(\{\b{h}\b{s}_i\}; \{\ell_{j}\})+P\prod_{j}\ell_{j}\Big),
\end{gather}
where $\b{h}=\text{diag}(\ell_1,\ell_2,\ell_3)$. One only has to construct the equations of motions to generate the equilibrium distribution (below $d\b{s}:=\prod_{i=1}^N d\b{s}_i$, and similar notations have been adopted for $d\b{p}^s$, $d\b{r}$ and $d\b{p}$)
\begin{gather}
	\exp(-\beta \mathscr{H}) d\b{s} d\b{p}^s d\ell_{j} dp_{j}^{\ell}
	=\exp\left[-\beta\left(\sum_i \frac{\b{p}_i^2}{2m_i}+\sum_{\alpha}\frac{|p_{j}^{\ell}|^2}{2M_{j}}+U+P\prod_j \ell_j \right)\right]d\b{r} d\b{p} d\ell_j dp_{j}^{\ell},
\end{gather}
which is the correct NPT ensemble.

Again, as for the cases with fixed shape, the partition function under the canonical ensemble can be written as an integral against $d\b{s}_i d\b{p}_i^s$ so that the dependence on $\b{h}$ can be explicit:
\begin{equation}
	Q(N, \b{h}, T)=\int \exp\left(-\beta\Big(\sum_i \frac{|\b{h}^{-1}\b{p}^s_i|^2}{2m_i}+U(\{\b{h}\b{s}_i\}; \{\ell_{j}\}) \Big) \right) d\b{s} d\b{p}^s.
\end{equation}
It is found that the pressure tensor is a diagonal matrix with the diagonal entry being
\begin{equation}
	\pi_{jj}=\frac{1}{V\beta}\frac{\partial \log Q}{\partial \ell_j}\ell_j=\int (P_{\mathrm{ins}})_{jj}\exp\left(-\beta\Big(\sum_i \frac{|\b{h}^{-1}\b{p}^s_i|^2}{2m_i}+U(\{\b{h}\b{s}_i\}; \{\ell_{\alpha}\}) \Big) \right) d\b{s} d\b{p}^s.
\end{equation}
Here, $V=\prod_{j}\ell_{j}$.  The instantaneous pressure is thus a diagonal matrix, with the diagonal entries being
\begin{gather}\label{e:instrectangular}
	(P_{\mathrm{ins}})_{jj}=\frac{1}{V}\frac{\partial (-K-U)}{\partial \ell_{j}}\ell_{j}
	=\frac{1}{V}\left[\sum_i\frac{(p_i)_{j}^2}{m_i}-\frac{\partial U(\{\b{h}\b{s}_i\}; \b{h})}{\partial \ell_{j}} \bigg{|}_{\b{s}_i\leftarrow \b{h}^{-1}\b{r}_i}\ell_{j}\right].
\end{gather}

Hence,
\begin{equation}
	\frac{\partial\mathscr{H}}{\partial \ell_j} =-V((P_{\mathrm{ins}})_{jj}-P)\ell_{j}^{-1}.
\end{equation}

The equations for the rectangular boxes can be given by
\begin{gather}
	\begin{split}
		&\dot{\b{r}}_i=\frac{\b{p}_i}{m_i}+\dot{\b{h}}\b{h}^{-1}\b{r}_i=\frac{\b{p}_i}{m_i}+\mathrm{diag}(\dot{\ell}_{j}/\ell_{j})\b{r}_i,\\
		&\dot{\b{p}}_i=-\nabla_{\b{r}_i}U-\mathrm{diag}(\dot{\ell}_{j}/\ell_{j})\b{p}_i-\gamma_i \b{p}_i+\sqrt{2k_BT\gamma m_i}\dot{\b{W}}_i,\\
		&\dot{\ell}_{j}=\frac{p^h_{j}}{M_{j}},\\
		& \dot{p}^{\ell}_{j}=V((P_{\mathrm{ins}})_{jj}-P)\ell_{j}^{-1}-\gamma_{j}p^{\ell}_{j}
		+\sqrt{2 k_BT M_{j}\gamma_{j}} \dot{W}_{j}.
	\end{split}
\end{gather}

It can be verified directly that the following virial theorems still hold
\begin{equation}
	\left\langle (P_{\text{ins}})_{jj}-P\right\rangle=\left\langle\frac{k_{\text{B}} T}{\operatorname{det}(\b{h})} \frac{\partial \log [Q(N,\b{h},T)]}{\partial \b{h}_{jj}} \b{h}_{jj}-P\right\rangle=0.
\end{equation}

\subsubsection*{Fully flexible cells}

The Lagrangian and Hamiltonian for the fully flexible cells have actually been discussed in \cite{parrinello1981polymorphic,gao2016sampling} already.
The only difference is that the potential in our periodic setting depends on $\b{h}$ explicitly.

To match the distribution function Eq.\eqref{flexible}, one may consider the Hamiltonian of the form
\begin{gather}\label{eq:Hamiltontianflexible}
	\mathcal{H}(\b{s}_i, \b{p}^s_i, \b{h}, \b{p}^h)=\sum_{i}\frac{|\b{h}^{-T}\b{p}_{i}^s|^2}{2m_i}
	+\sum_{\theta\eta}\frac{(p_{\theta\eta}^{h})^2}{2M_{\theta\eta}}
	+[U+P \det(\b{h})+\chi k_{\text{B}} T \ln[\det (\b{h})]].
\end{gather}
Here, $M_{\theta\eta}$ are some artificial parameters for the kinetic energy of the cells. This Hamiltonian \eqref{eq:Hamiltontianflexible} is the one in \cite{gao2016sampling}, which has an extra term $\chi k_B T \ln[\det (\b{h})]$
compared with the one in  \cite{parrinello1981polymorphic}.
As commented in Remark \ref{rmk:extratermfully}, the reason is that $d\b{h}$ is not the infinitesimal volume.

Again, the Jacobian of the transformation $\b{r}_i=\b{h}\b{s}_i, \b{p}_i=\b{h}^{-T}\b{p}^s_i$ is $1$, so that one has
\begin{equation}
	Q(N, \b{h}, T)=\int \exp\left[-\beta \left(\sum_{i}\frac{(\b{h}^{-T}\b{p}_{i}^s)^2}{2m_i}
	+U\right)\right] d\b{s} d\b{p}^s.
\end{equation}
Using the integration agains $\b{s}$ and $\b{p}^s$, the integral domain is then independent of $\b{h}$. Similarly as above, some calculation reveals that the pressure of the system is given by
\[
\pi_{\theta\eta}
=\frac{1}{(\det (\b{h}))\beta}\sum_{\gamma}\frac{\partial\log Q}{\partial h_{\theta\gamma}}h_{\eta\gamma}=\langle (P_{\mathrm{ins}})_{\theta\eta}\rangle_{(N,\b{h},T)},
\]
where the average is taken with respect to the canonical ensemble with the partition function $Q(N, \b{h}, T)$.  Defining $K(\b{h}, \b{p}^s_i)=\sum_{i}\frac{(\b{h}^{-T}\b{p}_{i}^s)^2}{2m_i}$ the kinetic energy, then one may  obtain
\begin{gather}\label{e:instflexible}
	\begin{split}
		(P_{\mathrm{ins}})_{\theta\eta} &=\frac{1}{\det (\b{h})} \sum_{\gamma}\frac{\partial (-K-U)}{\partial h_{\theta\gamma}}h_{\eta\gamma}\\
		&=\frac{1}{\det (\b{h})}\left(\sum_i \frac{\b{p}_i\otimes \b{p}_i }{m_i}- \sum_{\gamma}\frac{\partial U(\{\b{h}\b{s}_i\}, \b{h})}{\partial h_{\theta\gamma}}h_{\eta\gamma}\right) \Bigg|_{\b{s}_i\leftarrow \b{h}^{-1}\b{r}_i}.
	\end{split}
\end{gather}

\begin{remark}
	Comparing the expressions, the matrix in \eqref{e:instrectangular} is clearly the diagonal part of \eqref{e:instflexible}.
\end{remark}

Computing the partial derivative of $U$ on $h_{\theta\gamma}$ explicitly, we may read that the instantaneous pressure tensor is given by
\begin{gather}\label{eq:Pinsflexible}
	(P_{\mathrm{ins}})_{\theta\eta}
	=\frac{1}{\det (\b{h})}\left[\left(\sum_i \frac{\b{p}_i\otimes \b{p}_i }{m_i}-\nabla_{\b{r}_i}U \otimes \b{r}_i\right)_{\theta\eta}+\sum_{\gamma}-\frac{\partial U(\{\b{r}_i\}; \b{h})}{\partial h_{\theta\gamma}}\Bigg|_{\b{r}_i}h_{\eta\gamma}\right].
\end{gather}

The formula \eqref{eq:Pinsflexible} has one extra term
\begin{equation*}
	\sum_{\gamma}-\frac{\partial U(\{\b{r}_i\}; \b{h})}{\partial h_{\theta\gamma}}|_{\b{r}_i}h_{\eta\gamma},
\end{equation*}
as discussed in Remark \ref{rmk:extravirial}.
Again, for the Ewald summation, we would like to compute the derivatives of $U(\{\b{h}\b{s}_i\}; \b{h})$ with respect to $\b{h}$ by fixing $\b{s}_i$ directly, without computing this extra term.

When taking the average of $\pi_{\theta\eta}$ about the
NPT ensemble, or taking average with the weight to be $\exp(-\beta \det(\b{h}))(\det (\b{h}))^{1-\tilde{d}}d\b{h}$, one may verify the following virial theorem, as in \cite{martyna1994constant}
\begin{equation}
	\left\langle (P_{\text{ins}})_{\theta\eta}-P \delta_{\theta\eta}\right\rangle=\left\langle\frac{k_{\text{B}} T}{\operatorname{det}(\b{h})} \sum_{\gamma} \frac{\partial \log [Q(N,\b{h},T)]}{\partial h_{\theta\gamma}} h_{\eta\gamma}-P \delta_{\theta\eta}\right\rangle=0.
\end{equation}

The quantities $\frac{\partial \mathscr{H}}{\partial h_{\theta\eta}}$ can be written as
\begin{gather}
	\frac{\partial \mathscr{H}}{\partial h_{\theta\eta}}=-\det (\b{h})\left[\left(\b{P}_{\mathrm{ins}}-P -\frac{1}{\det (\b{h})}\chi k_{\text{B}}T\right) \b{h}^{-T}\right]_{\theta\eta},
\end{gather}
which will be used for constructing the equations of motions. The resulted equations for the fully flexible cells are given by
\begin{gather}
	\begin{split}
		& \dot{\b{r}}_i=\frac{\b{p}_i}{m_i}+\dot{\b{h}}\b{h}^{-1} \b{r}_i,~~~\dot{\b{p}}_i=-\nabla_{\b{r}_i}U-\b{h}^{-T}\dot{\b{h}}^T \b{p}_i-\gamma \b{p}_i+\sqrt{2\gamma k_{\text{B}}T m_i} \dot{\b{W}}_i,\\
		& \dot{h}_{\theta\eta}=\frac{p^h_{\theta\eta}}{M_{\theta\eta}},
		~~~\dot{p}^h_{\theta\eta}=\det(\b{h})\left[\left(\b{P}_{\mathrm{ins}}-P-\frac{\chi k_{\text{B}}T}{\det (\b{h})}\right)\b{h}^{-T}\right]_{\theta\eta}
		-\gamma_{\theta\eta} p^h_{\theta\eta}+\sqrt{2\gamma_{\theta\eta}k_{\text{B}}T M_{\theta\eta}} \dot{W}_{\theta\eta}.
	\end{split}
\end{gather}

\section{Hoover-type equations}\label{app:hoover}

We will also discuss the Hoover-type equations, using the Nos\'e-Hoover thermostats, to generate the desired ensembles given in section \ref{distributionfunction}.

We only present the equations for the simplest Hoover-type equations and the generalization to Nos\'e-Hoover chains should be straightforward. The idea of Nos\'e for generating the distribution $\exp(-\beta \mathcal{H}(p, q))$
is to consider an extended Hamiltonian
\begin{gather}
	\mathcal{H}_1(p', q', \zeta, p_{\zeta})=\mathcal{H}(p'/\zeta, q')+\frac{p_{\zeta}^2}{2M}+\frac{\Upsilon}{\beta}\ln \zeta.
\end{gather}
The partition function of the microcanonical ensemble for this Hamiltonian is given by
\begin{gather}
	\begin{split}
		\Delta(p', q', \zeta, p_{\zeta})&\propto \int \delta\left(\mathcal{H}(p'/\zeta, q')+\frac{p_{\zeta}^2}{2M}+\frac{\Upsilon}{\beta}\ln \zeta-E\right) dp' dq' d\zeta dp_{\zeta}\\
		&\propto \int \zeta^{1+\vartheta}\delta\left(\zeta-\exp\left(-\frac{\beta}{\Upsilon}\left[\mathcal{H}(p, q)+\frac{p_{\zeta}^2}{2M}-E\right]\right)\right) d\zeta dp dq dp_{\zeta}
	\end{split}
\end{gather}
so that the equilibrium distribution is given by $\exp\left(-\frac{(\vartheta+1) \beta}{\Upsilon}\mathcal{H}(p, q)\right) dp dq$, where $\vartheta$ is the dimension of $p$. Hence, one may run the following Hamilton ODEs
\begin{gather}
	\begin{split}
		&\frac{d}{d\tau}q'=\frac{1}{\zeta}\nabla_p\mathcal{H}(p'/\zeta, q'),~~\frac{d}{d\tau}p'=-\nabla_q\mathcal{H}(p'/\zeta, q'),\\
		&\frac{d}{d\tau}\zeta=\frac{p_{\zeta}}{M},~~~\frac{d}{d\tau}p_{\zeta}=\frac{p'}{\zeta^2}\cdot\nabla_p\mathcal{H}(p'/\zeta, q')-\frac{\Upsilon}{\beta \zeta}.
	\end{split}
\end{gather}
The idea of Hoover \cite{nose1984molecular,hoover1985canonical,hoover1986constant} is to consider a change of time $dt=d\tau/\zeta$, $p=p'/\zeta, q=q'$ so that the equations become
\begin{gather}
	\begin{split}
		&\frac{d}{dt}q=\nabla_p\mathcal{H}(p, q),~~\frac{d}{dt}p=-\nabla_q\mathcal{H}(p, q)-\xi p,\\
		&\frac{d}{dt}\xi=\frac{1}{M}\left(p\cdot\nabla_p\mathcal{H}(p, q)-\frac{\Upsilon}{\beta}\right),
	\end{split}
\end{gather}
where $\xi=d\zeta/dt$. However, under this uniform change of variables, the equilibrium distribution has been changed to
$\exp\left(-\frac{\vartheta \beta}{\Upsilon}\mathcal{H}(p, q)\right) dp dq$. Hence, one should use $\Upsilon=\vartheta$ for the correct distribution.
Note that the Hoover formulation is no longer a Hamilton ODE though it has an invariant quantity
\begin{equation}
	\mathcal{H}(p, q)+\dfrac{M\xi^2}{2}+\dfrac{\Upsilon}{\beta}\ln \zeta.
\end{equation}

Using the equations above for $\b{s}_i, \b{p}_i^s, V$ and changing back to the
variables $\b{r}_i, \b{p}_i, V$, one may find the equations of motion for the cubic boxes
\begin{gather}\label{e:newhoovercubic}
	\begin{split}
		& \dot{\b{r}}_i=\frac{\b{p}_i}{m_i}+\frac{\dot{V}}{3V} \b{r}_i,~~~\dot{\b{p}}_{i}= -\nabla_{\b{r}_i}U-\frac{\dot{V}}{3V}\b{p}_i
		-\xi \b{p}_i,\\
		& \dot{V}=\frac{p^V}{M},
		~~~\dot{p}^V=P_{\mathrm{ins}}-P -\xi p^V ,\\
		& \dot{\xi}=\frac{1}{M}\left[\sum_i \frac{|\b{p}_i|^2}{m_i}+\frac{(p^V)^2}{M}-\frac{3N+1}{\beta}\right].
	\end{split}
\end{gather}

This system of equations is very similar to the equations proposed by Hoover (see  \cite{hoover1985canonical,hoover1986constant}, and also \cite[Eq. (2.1)]{martyna1994constant}). However, the equation for the $p^V$, which corresponds to $3Vp^{\e}$ as in \cite[Eq. (2.1)]{martyna1994constant}, is different.
It is commented by the authors of \cite{martyna1994constant} that the equations proposed by Hoover do not generate the exact NPT distribution. Our equations \eqref{e:newhoovercubic} above, as derived from the first principles, can give the correct NPT ensemble. The MTK equations proposed in \cite{martyna1994constant} can also generate the correct NPT ensemble, but they achieved this through a correction term in Hoover's equations while the equations above are derived from the Hamiltonian directly and the equations for $p^V$ is different.

Similarly, for the fully flexible cells,  the Hoover-type equations are given by
\begin{gather}
	\begin{split}
		& \dot{\b{r}}_i=\frac{\b{p}_i}{m_i}+\dot{\b{h}}\b{h}^{-1} \b{r}_i,~~~\dot{\b{p}}_i=-\nabla_{\b{r}_i}U-\b{h}^{-T}\dot{\b{h}}^T \b{p}_i-\xi \b{p}_i,\\
		& \dot{h}_{\theta\eta}=\frac{p^h_{\theta\eta}}{M_{\theta\eta}},
		~~~\dot{p}_{\theta\eta}^h=\det(\b{h})[(\b{P}_{\mathrm{ins}}-P-\frac{\chi k_{\text{B}}T}{\det (\b{h})})\b{h}^{-T}]_{\theta\eta}
		-\xi p_{\theta\eta}^h,\\
		& \dot{\xi}=\frac{1}{M}\left[\sum_i \frac{|\b{p}_i|^2}{m_i}+\sum_{\theta\eta} \frac{(p^h_{\theta\eta})^2}{M_{\theta\eta}}-\frac{3N+9}{\beta}\right]
	\end{split}
\end{gather}
For rectangular boxes, one has
\[
\begin{split}
	&\dot{\b{r}}_i=\frac{\b{p}_i}{m_i}+\mathrm{diag}(\dot{\ell}_{j}/\ell_{j})\b{r}_i,
	\dot{\b{p}}_i=-\nabla_{\b{r}_i}U-\mathrm{diag}(\dot{\ell}_{j}/\ell_{j})\b{p}_i-\xi \b{p}_i,\\
	& \dot{\ell}_{j}=p_{j}^h,~~~\dot{p}^{\ell}_{j}=V((P_{\mathrm{ins}})_{jj}-P)\ell_{j}^{-1}-\xi p^{\ell}_{j},\\
	& \dot{\xi}=\frac{1}{M}\left[\sum_i \frac{|\b{p}_i|^2}{m_i}+\sum_{j} \frac{(p^h_{j})^2}{M_{j}}-\frac{3N+3}{\beta}\right].
\end{split}
\]

\begin{remark}
	In all these equations, using the same $\xi$ for all variables might not be very suitable, as characteristic times of the thermostat of particles and the box might be different.
	The Langevin equations in sections \ref{equationofmotion} and \ref{extensions} allow different friction parameters for different variables. For these Hoover type equations, one may consider (several) Nos\'e-Hoover chains to separate the time scales so that one may use different scaling parameters $\xi$ for the particles and the cell as discussed in \cite[section E]{martyna1994constant}.
\end{remark}

\section{The random batch Ewald method for general cases }\label{app:rbegeneral}

In practice, two particular cases may be: the cell is allowed to rotate, or the angles in the simulation cell may not right angles, i.e., the cell becomes triclinic. The Ewald sum and the random mini-batch approach need some modification. In this subsection, we first recall the Ewald sum in the general setting and then discuss how the random batch Ewald strategy may be applied for these cases.

We define the reciprocal vectors
\begin{equation}
	\b{b}_1=2\pi \frac{\b{h}_2\times \b{h}_3}{\Omega},~~\b{b}_2=2\pi \frac{\b{h}_3\times \b{h}_1}{\Omega},~~\text{and}~\b{b}_3=2\pi \frac{\b{h}_1\times \b{h}_2}{\Omega},
\end{equation}
with $\Omega=\det(\b{h})$. Then, the relations $\b{h}_i\cdot \b{b}_j=2\pi \delta_{ij}$ can be established for all $i,j\in\{1,2,3\}$. The wavenumber is then given by
\begin{equation}
	\b{k}=\sum_i m_i \b{b}_i,~~m_i\in\mathbb{Z}.
\end{equation}
In other words, $[\b{b}_1, \b{b}_2, \b{b}_3]=2\pi \b{h}^{-T}$ where the superscript $T$ indicates the transpose, and $\b{k}=2\pi \b{h}^{-T}\b{m}$ with $\b{m}=[m_1,m_2,m_3]^T$.

Consider the two parts, $U_1$ and $U_2$, of the Coulomb energy given in Eq.\eqref{eq:U1fourier} and Eq.\eqref{U2}, respectively. Using the Fourier transform
$\tilde{f}(\b{k})=\int_{\b{h}} f(\b{r}) e^{-i \b{k}\cdot \b{r}} d\b{r}$,
and its inverse transform $\frac{1}{\det (\b{h})}\sum_{\b{k}} \tilde{f}(\b{k})e^{i\b{k} \cdot \b{r}}$, it is easy to find that for the general cells, the expressions of $U_1$ and $U_2$ are still given by
\begin{equation}
	U_1(\{\b{h}\b{s}_i\}; \b{h})=\frac{2\pi}{\det (\b{h})} \sum_{\b{k}\neq \b{0}}\frac{1}{|\b{k}|^2} \left|\sum_i q_i e^{i \b{k}\cdot \b{r}_i}\right|^2\Bigg{|}_{\b{r}_i=\b{h}\b{s}_i}
	\exp(-|\b{k}|^2/(4\alpha)),
\end{equation}
and
\begin{equation}
	U_2=\frac{1}{2}\sum_{\b{n}}{'}\sum_{ij}  q_i   q_j \frac{1}{|\b{r}_{ij}+\b{h}\b{n}|}
	\mathrm{erfc}\left(\sqrt{\alpha}|\b{r}_{ij}+\b{h}\b{n}|\right).
\end{equation}
The forces can be computed the same as before and given by \eqref{eq:force}.
For the pressure computation, we also note that $\b{k}\cdot \b{r}_i$ is in fact independent of $\b{h}$ and $\b{r}_{ij}+\b{h}\b{n}
=\b{h}(\b{s}_i-\b{s}_j+\b{n})$, the expressions can be derived totally similarly as in section \ref{Ewaldsum}. The resulted expression for the fully flexible cells is given by
\begin{gather}\label{eq:pressuretensorflexible}
	\begin{split}
		(P_{\text{ins}})_{\theta\eta}
		&=\frac{1}{\det \b{h}}\sum_i \frac{(\b{p}_i\otimes \b{p}_i)_{\theta\eta} }{m_i}+\frac{1}{\det (\b{h})}\sum_{\gamma}\frac{\partial(-U)}{\partial \b{h}_{\theta\gamma}}\b{h}_{\eta\gamma}\\
		&=\frac{1}{\det (\b{h})}\sum_i \frac{(\b{p}_i\otimes \b{p}_i)_{\theta\eta} }{m_i}+\frac{1}{\det (\b{h})}\left[\frac{1}{2}\sum_{\b{n}}{'}\sum_{ij}  \b{F}_{ij,\b{n}}\otimes (\b{r}_{ij}+\b{h}\b{n}) \right]_{\theta\eta}\\
		&+\frac{2\pi}{\det (\b{h})^2} \sum_{\b{k}\neq \b{0}}\frac{1}{|\b{k}|^2}|\rho(\b{k})|^2\exp(-|\b{k}|^2/(4\alpha))
		\left[\b{I}-2\left(\frac{1}{\b{k}^2}+\frac{1}{4\alpha}\right)\b{k}\otimes \b{k}\right]_{\theta\eta},
	\end{split}
\end{gather}
where $\theta,\eta,\gamma\in\{1,2,3\}$, $\b{I}$ is the three-dimensional identity matrix, and
\begin{equation}
	\b{F}_{ij,\b{n}}=-q_iq_j G(|\bm{r}_{ij}+\b{h}\bm{n}|)\frac{\bm{r}_{ij}+\b{h}\bm{n}}{|\bm{r}_{ij}+\b{h}\bm{n}|}.
\end{equation}

The rectangular case is similar. In fact, the pressure tensor, as has been seen,
is a diagonal matrix. Recall $V=\prod_{\theta=1}^3\ell_{\theta}$
and $k_{\theta}=2\pi m_{\theta}/\ell_{\theta}$, and $\b{h}=\mathrm{diag}(\ell_{1}, \ell_{2},\ell_{3})$.
Direct computation reveals that
\begin{gather}\label{eq:pressurerect}
	\begin{split}
		(P_{\text{ins}})_{\theta\theta}&=\frac{1}{V}\sum_i\frac{(\b{p}_i)_{\theta}^2}{m_i}
		+\frac{1}{2V}\sum_{\b{n}}{'}\sum_{ij}  (F_{ij,\b{n}})_{\theta}(\b{r}_{ij}+\b{h}\b{n})_{\theta}\\
		&+\frac{2\pi}{V^2}\sum_{\b{k}\neq \b{0}}\frac{1}{|\b{k}|^2}|\rho(\b{k})|^2\exp(-|\b{k}|^2/(4\alpha))
		\left[1-2\left(\frac{1}{|\b{k}|^2}+\frac{1}{4\alpha}\right)k_{\theta}^2\right],
	\end{split}
\end{gather}
and $(P_{\text{ins}})_{\theta\eta}=0$ for $\theta\neq\eta$. Clearly, the frequency part here agrees with the one in \cite{brown1995general}.

\begin{remark}
	Clearly, the pressure \eqref{eq:pressurecubic} in the isotropic case equals $1/3$ of the trace of the tensors \eqref{eq:pressuretensorflexible} and \eqref{eq:pressurerect} above.
\end{remark}

Below, we make some remarks about sampling of the frequencies in the random batch Ewald sum for rectangular and full flexible cells:
\begin{enumerate}
	\item
	For rectangular $\b{h}$, the sampling is similar to section \ref{sample} except that the three components of $\b{m}$ satisfy different discrete Gaussians, but they are still independent. There is no big difference.
	
	\item
	It becomes tricky for general cell shapes, where the integer vector
	$\b{m}$ now obeys
	\begin{equation}
		\mathscr{P}(\b{m})\propto \exp(-\pi^2\b{m}^T(\b{h}^{T}\b{h})^{-1}\b{m}/\alpha),~~\b{m}\neq 0.
	\end{equation}
	To sample from this distribution, we may use the approximate offline approach: we simply sample from the continous distribution $\exp(-\pi^2\b{m}^T(\b{h}^{T}\b{h})^{-1}\b{m}/\alpha)$
	and convert it to an integer vector.
	For the exact sampling, using the continuous Gaussian as proposal seems troublesome as the probability that the continous Gaussian falls into $\otimes_{\theta=1}^3 I_{m_\theta}$ where $I_{\theta}=[m_{\theta}-1/2, m_{\theta}+1/2)$ cannot be computed easily. One may run the MH algorithm on the discrete points using discrete proposals directly, which of course will cost more, whereas this cost could be reduced by using the third strategy provided in section \ref{sample}.
\end{enumerate}

\bibliographystyle{plain}
\bibliography{NPT}

\end{document}